\documentclass[aps,twocolumn,showpacs]{revtex4}
\usepackage{graphicx} 
\usepackage{caption}  
\usepackage{subcaption}
\usepackage{amsmath,color,ulem}
\usepackage{physics}
\usepackage{epsfig}
\usepackage{multirow}
\usepackage{float}
\usepackage{array} 
\usepackage{rotating}

\usepackage{booktabs} 

\begin{document}
	
	\title{Investigation of deuteron-like singly bottomed dibaryon resonances}
\author{Yuxuan Du$^1$}
\author{Yanyue Pan$^1$}
\author{Xinmei Zhu$^2$}
\author{Zhiyun Tan$^3$}
\author{Hongxia Huang$^1$}\email[E-mail: ]{hxhuang@njnu.edu.cn (Corresponding author)}
\author{Jialun Ping$^1$}
\affiliation{$^1$Department of Physics, Nanjing Normal University, Nanjing, Jiangsu 210097, China}
\affiliation{$^2$Department of Physics, Yangzhou University, Yangzhou 225009, People's Republic of China}
\affiliation{$^3$School of Physics and Electronic Science, Zunyi Normal College, Zunyi, Guizhou 563006, China}

	\begin{abstract}
		
		We perform a systematical investigation of the existence of the deuteron-like singly bottomed dibaryon resonance states with strangeness $S=-1,~-3,~-5$ in the chiral quark model.
		Two resonance states with strangeness $S=-1$ are obtained in the baryon-baryon scattering process.
		The first candidate is $\Sigma\Sigma_b$ in the $\Lambda\Lambda_b$ and $N\Xi_b^*$ scattering process, with the resonance energy 6974.22 MeV - 6975.37 MeV and the decay width 14.450 MeV, respectively; the other one is $\Sigma \Sigma_b^*$ in the $N\Xi_b$ and $N\Xi'_b$ scattering process, with the resonance energy 6990.69 MeV - 7008.37 MeV and the decay width 43.790 MeV, respectively. The Root Mean Square (RMS) radius calculation shows that the former tends to be in a compact structure, while the latter tends to be in a molecular structure. Both of these resonance states are worthy of experimental exploration. Furthermore, it should be emphasized that the effect of channel-coupling is of great importance in exploring exotic hadron states, and investigating the scattering process may serve as an effective approach to identifying genuine resonances.

	\end{abstract}

	\maketitle
	
	\setcounter{totalnumber}{5}
	
	\section{\label{sec:introduction}Introduction}
	
	 Dibaryon states represent an important research topic for exploring the property of quantum chromodynamics (QCD). The deuteron, discovered in 1932 as a bound state of a proton and a neutron, is currently the only known stable dibaryon system \cite{HCU}. In 1977, based on the MIT
bag model, Jaffe employed gluon exchange interaction to identify a stable six-quark configuration: H-dibaryon \cite{H}. Over the past decade, researchers have persisted in searching for new dibaryon states. The discovery of the dibaryon resonance state $d^*$ has undoubtedly attracted the attention of many experimentalists and theorists \cite{TGKM}-\cite{PACAW}. Physicists have also investigated this dibaryon state using various theoretical approaches \cite{JLP}-\cite{HXCE}. Regarding the dibaryon states with strange quarks, significant attention has been focused on the H-dibaryon, $N\Omega$, and $\Omega\Omega$ states. Here, we will take $N\Omega$ as a representative example for discussion.
In 1987, T. Goldman, using two different quark models, proposed that a dibaryon state with the strangeness $S=-3$ might exist \cite{JTG}. Subsequently, researchers have employed various theoretical approaches to confirm the existence of $N\Omega$, such as quark models \cite{omf}-\cite{dze} and lattice QCD \cite{FEHN}-\cite{TIS}. Experiments have also confirmed the existence of the $N\Omega$ state. In 2019, STAR collaboration presented the first measurement of the $p$–$\Omega$ correlation function in $Au+Au$ collisions at $\sqrt{s_{NN}}=200$ GeV, offering supporting evidence for the existence of $p$–$\Omega$ with a binding energy of 27 MeV \cite{aj}. In 2020, the ALICE collaboration showed that $p$–$\Omega$ interaction can be investigated with remarkable precision, which the signal was up to a factor two larger than the $p$–$\Xi$ signal \cite{ac}.

The extensive studies on multiquark states with light quarks have prompted researchers to consider the existence of exotic states containing heavy quarks.
In 1981, the CERN collaboration first detected a singly bottom baryon state, named it $\Lambda_{b}^0$ \cite{mb}. Fourteen years later, the DELPHI detector observed the first singly strange bottom baryon states: $\Xi_{b}^0$ and $\Xi_{b}^{-}$ \cite{pa}. Subsequently, in 2007, the CDF experiment reported the discoveries of $\Sigma_{b}^{\pm}$, $\Sigma_{b}^{*\pm}$ \cite{ta}. The discoveries of baryons with a bottom quark sparked significant interest among researchers in exploring bottomed dibaryon states. Theoretically, the relatively large mass of the bottom quark suppresses the kinetic term in the Hamiltonian, thereby enhancing the feasibility of bound states.~As a result, dibaryon systems containing bottom quarks have attracted growing research interest.~Researchers have adopted multiple theoretical approaches to study bottomed dibaryons. Some theoretical studies have employed the lattice QCD to investigate dibaryon states involving bottom quarks \cite{pj}-\cite{pmn}. In Ref. \cite{pj}, P. Junnarkar et al.~unambiguously found that the ground state masses of dibaryons $\Omega_{b}\Omega_{bb}(ssbsbb),~\Omega_{ccb}\Omega_{cbb}(ccbcbb)$ were below their respective two baryon thresholds, suggesting the presence of bound states that were stable under strong and electromagnetic interactions.~Some theoretical studies have employed the one boson exchange model to investigate heavy dibaryon states \cite{mzl}. In Ref.~\cite{mzl}, M. Z. Liu et al. extended the conventional one boson exchange model by allowing for exchanges of heavy $s\bar{s}$, $b\bar{b}$ ground state mesons.~They predicted the existence of one $J^{P}=0^{+}$ $\Omega_{bbb}\Omega_{bbb}$ bound state with considerably small about 6 MeV.
In Ref.~\cite{jjq}, J. J. Qi et al. systematically studied the heavy baryonium and heavy dibaryon systems using the Bethe-Salpeter equation in the ladder and instantaneous approximations for the kernel. Their results indicated that all the heavy baryonium systems, specifically $\Xi_{b}{\bar{\Xi}_{b}}$, $\Lambda_{b}{\bar{\Lambda}_{b}}$, $\Sigma_{b}{\bar{\Sigma}_{b}}$, $\Xi^{'}_{b}{\bar{\Xi}^{'}_{b}}$ and $\Omega_{b}{\bar{\Omega}_{b}}$, can form bound states. Among the heavy dibaryon systems, only the  $\Xi_{b}\Xi_{b}$ system with $I=0$ and the $\Sigma_{b}\Sigma_{b}$ systems with $I=0$ and $I=1$ can exist as bound states.
Some theoretical groups investigated heavy dibaryons within the framework of the quark model \cite{db}-\cite{pmh}. In Ref. \cite{jma}, J. M. Alcalaz-Pelegrina et al. used a diffusion Monte Carlo technique to describe the properties of fully heavy compact six-quark systems within the framework of the constituent quark model. In all cases, the masses of the six-quark systems were larger than those corresponding to the sum of any two baryons but smaller than those for a set of six isolated quarks. That was, all of them were bound systems. In Ref.~\cite{pmh}, P. M. Higueras et al. studied the possible existence of fully heavy dibaryons in the bottom sectors. The binding energy of the bottom case was $-1.98$ MeV, which was reasonable due to the highest mass of the bottom quark.
In our group, we also employ the quark model to investigate dibaryons containing heavy quarks \cite{hxh}-\cite{zcx}. In Ref. \cite{hxh}, H. X. Huang et al. investigated possible H-like dibaryon states $\Lambda_{b}\Lambda_{b}$ within Quark Delocalization Color Screening Model (QDCSM), the results showed that bound state was also possible in the $\Lambda_{b}\Lambda_{b}$ system due to the central interaction of one-gluon exchange and one-pion exchange. In Ref.~\cite{zcx}, X. Z. Cheng et al. performed a systemical investigation of the low-lying doubly heavy dibaryon systems with strange $S=0$ in QDCSM. Three resonance states were obtained, which were $N\Xi_{bb}$, $N\Xi^*_{bb}$, $\Sigma_{b}\Sigma^*_{b}$. All these heavy dibaryons are worth searching for in experiments.

The systematic search for heavy dibaryon states is an intricate task.~Therefore, in this work, we have conducted a systematic investigation of deuteron-like dibaryons containing a singly bottomed quark in Chiral Quark Model (ChQM), with strangeness $S=-1,~-3,~-5$, angular momentum quantum number $J$ = 1, and isospin $I=0$. The structure of this paper is organized as follows: Section \ref{Quark model and resonating group method} provides a introduction to the quark model and calculation methods. Section \ref{11} provides the numerical results and corresponding discussion. Section \ref{22} concludes the paper and the last is Appendix, which shows the Gell-Mann matrices and the construction of the wave function.

\section{\label{Quark model and resonating group method}Quark model and calculation methods}
This work is conducted within the framework of the chiral quark model (ChQM). In order to search for bound states and resonance states, we use the Resonating Group Method (RGM) \cite{RGM1}-\cite{RGM} and Kohn-Hulthen-Kato (KHK) variational method \cite{MKA}-\cite{mot}, respectively.
This chapter will provide a brief introduction to the model and the methods.
	\subsection{The chiral quark  model}
	The chiral quark model is one of the most common approaches to describe hadron spectra, hadron-hadron interactions and multiquark states \cite{AVHG}. In this model, the short range interaction is primarily provided by the one-gluon exchange potential, the intermediate range attraction is generated by the $\sigma$ meson exchange potential, while the Goldstone boson exchange potential dominates the long range interaction.  Here we only show the Hamiltonian and the parameters used.
\begin{equation}
	H=\sum_{i=1}^6\left(m_i+\frac{p_i^2}{2m_i}\right)-T_c+\sum_{i<j}V_{ij}
	\label{eq:Hamiltonian}
\end{equation}
	\begin{equation}
		V_{ij} = V^{\text{CON}}(r_{ij})+V^{\text{OGE}}(r_{ij}) + V^{\sigma}(r_{ij})
		+V^{\text{OBE}}(r_{ij}) \\
	\end{equation}
		Where $m_i$ is the mass of different quarks, the kinetic energy term in the Hamiltonian is $\frac{p_i^2}{2m_i}$, the term for the kinetic energy of the center of mass is $T_c$. The potential interaction $V_{ij}$ includes the confinement potential $V^{\text{CON}}(r_{ij})$, the one-gluon exchange potential $V^{\text{OGE}}(r_{ij})$, the $\sigma$ meson exchange potential $V^{\sigma}(r_{ij})$, and the one-boson exchange potential $V^{\text{OBE}}(r_{ij})$.~Here, we will present the expressions for each potential. The confinement potential $V^{\text{CON}}(r_{ij})$ is displayed:
	\begin{equation}
		V^{\text{CON}}(r_{ij})= -a_c \boldsymbol{\lambda_i} \cdot \boldsymbol{\lambda_j} \left[ r_{ij}^2 + V_0 \right] \label{eq:VCON} 
	\end{equation}
	Additionally, based on the asymptotic freedom property of QCD, the model introduces the one-gluon exchange potential:
	\begin{align}
		V^{\text{OGE}}(r_{ij}) &= \frac{1}{4} \alpha_{s_{q_{i}q_{j}}} \boldsymbol{\lambda_i} \cdot \boldsymbol{\lambda_j} \Bigg[ \frac{1}{r_{ij}}  	- \frac{\pi}{2} \delta(\boldsymbol{r_{ij}})  \nonumber \\
		         & \left( \frac{1}{m_i^2} + \frac{1}{m_j^2} + \frac{4\boldsymbol{\sigma}_i \cdot \boldsymbol{\sigma}_j}{3m_i m_j} \right) - \frac{3}{4m_i m_j r_{ij}^3} S_{ij} \Bigg] \\ \label{eq:VOGE}
		         S_{ij} &= \frac{(\boldsymbol{\sigma}_i \cdot \boldsymbol{r}_{ij})(\boldsymbol{\sigma}_j \cdot \boldsymbol{r}_{ij})}{r_{ij}^2} - \frac{1}{3} \boldsymbol{\sigma}_i \cdot \boldsymbol{\sigma}_j
	\end{align}
   Where $S_{ij}$ is the quark tensor operator. We consider only $S$-wave states in this work, the tensor force operator does not contribute. In the chiral quark model, we introduce the scalar $\sigma$ meson exchange potential (only between $u$ and $d$ quarks) to provide intermediate range attraction. Its specific expression is:
	\begin{align}
		V^{\sigma}(r_{ij}) &= -\frac{g^{2}_{\text{ch}}}{4\pi}\frac{\Lambda_{\sigma}^{2}m_{\sigma}}{\Lambda_{\sigma}^{2}-m_{\sigma}^{2}}
		\left[Y(m_\sigma r_{ij}) - \frac{\Lambda_{\sigma}}{m_{\sigma}}Y(\Lambda_{\sigma} r_{ij})\right]  \label{eq:Vsigma}
		\end{align}\\
		
		In the standard framework, the Goldstone boson exchange potential is composed solely of $\pi$, $K$, $\eta$ meson exchanges, but this study focuses on dibaryons with a bottom quark, necessitating an extension from $SU(3)$ to $SU(5)$ flavor symmetry. Within this expanded framework, the standard Goldstone boson exchange potential is no longer sufficient, we have incorporated the following meson exchange potentials: $B$-meson exchange potential between $u/d$ and $b$ quarks, $B_s$-meson exchange potential between $s$ and $b$ quarks, $\eta_b$-meson exchange potential between any two quarks ($u/d$, $s$ or $b$ quarks).
		Therefore, the specific expression for $V^{\text{OBE}}(r_{ij})$ is expressed:
		\begin{align}
		V^{\text{OBE}}(r_{ij}) &=
		\nu_{\pi}(r_{ij})\sum_{a=1}^3 \lambda^{a}_i \cdot \lambda^{a}_j
		+ \nu_{K}(r_{ij})\sum_{a=4}^7 \lambda^{a}_i \cdot \lambda^{a}_j \nonumber \\
		&\quad + \nu_{\eta}(r_{ij})\left[
		\left(\lambda^{8}_i \cdot \lambda^{8}_j\right)\cos\theta_p
		- \left(\lambda^{0}_i \cdot \lambda^{0}_j\right)\sin\theta_p
		\right] \nonumber \\
		&\quad + \nu_{B}(r_{ij})\sum_{a=16}^{19} \lambda^{a}_i \cdot \lambda^{a}_j
		+ \nu_{B_s}(r_{ij})\sum_{a=20}^{21} \lambda^{a}_i \cdot \lambda^{a}_j \nonumber \\
		&\quad + \nu_{\eta_b}(r_{ij})\lambda^{24}_i \cdot \lambda^{24}_j \nonumber  \label{eq:VOBE} \\
		\end{align}
		\begin{align}
		\nu_{\chi}(r_{ij}) &= \frac{g^{2}_{\text{ch}}}{4\pi}\frac{m_{\chi}^2}{12m_{i}m_{j}}
		\frac{\Lambda_{\chi}^2}{\Lambda_{\chi}^2-m_{\chi}^2}m_\chi \Bigg\{
		\nonumber \\
		&\left[Y(m_{\chi} r_{ij}) - \frac{\Lambda_{\chi}^3}{m_{\chi}^3}Y(\Lambda_{\chi} r_{ij})\right] \boldsymbol{\sigma}_i \cdot \boldsymbol{\sigma}_j \nonumber \\
		&\quad + \left[H(m_{\chi} r_{ij}) - \frac{\Lambda_{\chi}^3}{m_\chi^3} H(\Lambda_{\chi} r_{ij})\right] S_{ij} \Bigg\} \label{eq:nux} 	
	\end{align}
	In Eq (\ref{eq:VOBE}), the symbol $\lambda^{a}$ represents Gell-Mann matrices, which are presented in appendix \ref{Gell-Mann matrices}.
In Eq (\ref{eq:nux}), $Y (x)$ and $H(x)$ are standard Yukawa functions \cite{AVHG}.
The model parameters are determined as follows. The mass parameters (e.g. $m_\pi$, $m_K$, $m_\eta$, $m_B$, $m_{B_{s}}$, $m_{\eta_{b}}$) take their experimental values. The mass of $u/d$-quark ($m_{u,d}$), cutoff parameters (e.g. $\Lambda_\pi$, $\Lambda_K$, $\Lambda_\eta$) and the mixing angle $\theta_p$ are empirically fit~\cite{VIJ}. The chiral coupling constant $g_{ch}$ can be obtained from the $\pi NN$ coupling constant through
\begin{align}
\frac{g^{2}_{ch}}{4\pi}= \left(\frac{3}{5}\right)^{2}\frac{g^{2}_{\pi NN}}{4\pi}\frac{m^{2}_{u,d}}{m^{2}_{N}}.
       \end{align}
The other adjustable parameters can be determined by fitting the ground-state light baryons and singly heavy baryons. In quark model calculations, equivalent coupling constants are used to describe the spectra of baryons and mesons. As shown in $V^{\text{OGE}}$, the chromomagnetic term is related to $\frac{\alpha_{s_{q_{i}q_{j}}}}{m_i m_j}$, where $\alpha_{s_{q_{i}q_{j}}}$ is associated with the quark flavor and determined by the mass difference between two baryons with spins $S=\frac{1}{2},~S=\frac{3}{2}$, respectively. The mass of heavy quarks is relatively large, which reduces the mass difference between the two baryons with spins $S=\frac{1}{2},~S=\frac{3}{2}$. This makes it necessary to increase the value of $\alpha_{s_{q_{i}q_{j}}}$ to counteract the effect of quark mass and thereby obtain a mass difference that aligns more closely with experimental measurements.
The model parameters and the masses of the fitted baryons are shown in
Table~\ref{Model parameters} and Table~\ref{masses}, respectively. The energies listed in Tables \ref{tab1}-\ref{tab2} are expressed in the natural unit system ($\hbar$ = $c$ = 1).
	
\begin{table}[H]
		\centering
		\caption{Model parameters: $m_{\pi}=0.70~fm^{-1}$, $m_{K}=2.51~fm^{-1}$, $m_{\eta}=2.77~fm^{-1}$, $m_{\sigma}=3.42~fm^{-1}$, $m_{B}=26.75~fm^{-1}$, $m_{B_s}=27.20~fm^{-1}$, $m_{\eta_b}=47.63~fm^{-1}$, $\Lambda_{\pi}=\Lambda_{\sigma}=4.20~fm^{-1}$, $\Lambda_{K}=\Lambda_{\eta}=5.20~fm^{-1}$, $\Lambda_{B}=\Lambda_{B_s}=\Lambda_{\eta_b}=1.20~fm^{-1}$.}
		\label{Model parameters}
		\begin{tabular}{c*{9}{c}}  
			\hline
			\hline
			& $b(fm)$ & $m_{u,d}($MeV$)$ & $m_{s}($MeV$)$ & $m_{c}($MeV$)$ & $m_{b}($MeV$)$  \\
			
			 & 0.49088 & 313 & 590 & 1700 & 5244.1 \\
			\hline
			 & $a_{c}($MeV$ \cdot fm^{-2})$  & $V_0(fm^{2})$ & $\alpha_{s_{qq}}$ & $\alpha_{s_{qs}}$ & $\alpha_{s_{ss}}$  \\
			  & 50.330  & -1.2779 & 0.50594 & 0.82961 & 0.74093 \\
			\hline
			  & $\alpha_{s_{qb}}$& $\alpha_{s_{sb}}$ \\
			 & 0.72087&0.99010\\
          \hline			
		\hline	
		\end{tabular}
	\end{table}
	
	\begin{table}[H]
		\centering
		\caption{The calculated masses (in MeV) of the baryons in
ChQM. Experimental values are taken from the Particle Data
Group (PDG) \cite{JB}.}
		\label{masses}
		\begin{tabular}{c*{9}{c}}  
			\hline
			\hline
			& $N$ & $\Delta$ & $\Lambda$ & $\Sigma$ & $\Sigma^*$ & $\Omega$ & $\Xi$ & $\Xi^*$ & $\Lambda_b$  \\
			\hline
			\text{ChQM} & 944 & 1276 & 1067 & 1148 & 1345 & 1564 & 1249 & 1446 & 5646 \\
			\text{Exp.} & 939 & 1233 & 1116 & 1189 & 1315 & 1672 & 1385 & 1530 & 5620 \\
			\hline
			
			\hline
			& $\Sigma_b$  & $\Sigma^*_b$ & $\Xi_b$ & $\Xi'_b$ & $\Xi^*_b$  & $\Omega_b$ \\
			\hline
			\text{ChQM} & 5848 & 5863 & 5769 & 5892 & 5905 & 5953 \\
			\text{Exp.} & 5811 & 5830 & 5792 & 5935 & 5945 & 6046 \\
			\hline
			\hline
		\end{tabular}
	\end{table}

	\subsection{\label{The Resonating Group Method}The caculation methods}
We employ the RGM to conduct a dynamical calculation. For a bound state problem, we express the wave function of the baryon - baryon system as: \begin{align}
		\Psi_{6q} = {\cal A }  \left[[\phi_{B_1}\left(\boldsymbol{\rho_1},\boldsymbol{\lambda_1}\right)
		{\phi}_{B_2}\left(\boldsymbol{\rho_2},\boldsymbol{\lambda_2}\right)]^{[\sigma]IS}\chi(\boldsymbol{R})Z\left(\boldsymbol{R_c}\right)\right]^{J}
		\label{phi5q}
	\end{align}
	where the symbol ${\cal A }$ is the anti-symmetrization operator. With the $SU(5)$ extension, both the light and heavy quarks are considered as identical particles, so the symbol ${\cal A }$ is written as ${\cal A }=1-9P_{36}$. Meanwhile, $\sigma=[222]$ gives the total color symmetry, $I$ represents isospin quantum number and $S$ refers to the spin quantum number. $Z\left(\boldsymbol{R_c}\right)$ denotes the center of mass motion wave function of the two clusters, where $\boldsymbol{R_c}$ represents the center-of-mass coordinate.  The internal wave functions of the two three-quark clusters is ${\phi}_{B_i}$, both $\boldsymbol{\rho_i}$ and $\boldsymbol{\lambda_i}$ are internal coordinates. We assume that the wave function ${\phi}_{B_i}$ is the product of the harmonic oscillator ground state wave function and the isospin-spin wave function $\phi_{I_i S_i}\left(B_i\right)$ and color wave function $\chi_{c_i}$, expressed as:
	\begin{align}
	\phi_{B_i} = \left( \frac{2}{3\pi b^2} \right)^{3/4} &\left( \frac{2}{4\pi b^2} \right)^{3/4} \exp\left( -\frac{\boldsymbol{\lambda_i}^2}{3 b^2}-\frac{\boldsymbol{\rho_i}^2}{4 b^2} \right) \nonumber\\
	& \times\phi_{I_i S_i}\left(B_i\right) \chi_{c_i}\left(B_i\right)
	\end{align}
	The relative motion wave function between the two clusters is $\chi(\boldsymbol{R})=\sum_{L}\chi_{L}(\boldsymbol{R})$, in which $\boldsymbol{R}$ refers to the relative coordinate and $L$ is the orbital angular momentum between the two clusters.
	 Now we can determine the relative motion wave function by solving the Schrödinger equation. From the variational principle:
	\begin{align}
	\langle \delta \Psi'' | H - E | \Psi' \rangle = 0
     \end{align}	
	  After performing the variation, we can obtain the RGM equation:
	\begin{align}
		\int H(\boldsymbol{R''},\boldsymbol{R'})\chi(\boldsymbol{R'})\,d\boldsymbol{R'} = E \int N(\boldsymbol{R''},\boldsymbol{R'})\chi(\boldsymbol{R'})\,d\boldsymbol{R'}
		\label{phi6q}
	\end{align}
	where $H(\boldsymbol{R''},\boldsymbol{R'})$ and $N(\boldsymbol{R''},\boldsymbol{R'})$ are Hamiltonian and norm kernels, specifically expressed as:
	
	\begin{align}
	H(\boldsymbol{R''}, \boldsymbol{R'}) =H^D\left(\boldsymbol{R''},\boldsymbol{R'}\right)  \delta(\boldsymbol{R} - \boldsymbol{R''}) +H^{EX} \left(\boldsymbol{R''},\boldsymbol{R'}\right)  \nonumber \\
N(\boldsymbol{R''}, \boldsymbol{R'}) =N^D\left(\boldsymbol{R''},\boldsymbol{R'}\right)  \delta(\boldsymbol{R} - \boldsymbol{R''}) +N^{EX} \left(\boldsymbol{R''},\boldsymbol{R'}\right)
     \end{align}
where the superscript $D$ represents the direct item, and the superscript $EX$ represents the exchange item.
 Then, integrate with respect to the coordinates $\boldsymbol{\rho_{1}}$, $\boldsymbol{\rho_{2}}$, $\boldsymbol{\lambda_{1}}$, $\boldsymbol{\lambda_{2}}$ and $\boldsymbol{R}$, we can obtain the formula:
	\begin{align}
        \begin{pmatrix}
            N^D(\boldsymbol{R''}, \boldsymbol{R'}) \\
            H^D(\boldsymbol{R''}, \boldsymbol{R'})
        \end{pmatrix} \delta(\boldsymbol{R''} - \boldsymbol{R'})
         = \int \phi_1^*(\boldsymbol{\rho_1}, \boldsymbol{\lambda_1}) \phi_2^*(\boldsymbol{\rho_2}, \boldsymbol{\lambda_2}) \nonumber \\
        \delta(\boldsymbol{R} - \boldsymbol{R''})
        \begin{pmatrix}
            1 \\
            H
        \end{pmatrix}
         \times \phi_1(\boldsymbol{\rho_1}, \boldsymbol{\lambda_1}) \phi_2(\boldsymbol{\rho_2}, \boldsymbol{\lambda_2}) \nonumber \\
         \delta(\boldsymbol{R} - \boldsymbol{R'})  \, d\boldsymbol{\rho_1} \, d\boldsymbol{\lambda_1} \, d\boldsymbol{\rho_2} \, d\boldsymbol{\lambda_2} \, d\boldsymbol{R}
       \end{align}
       \begin{align}
        \begin{pmatrix}
            N^{EX}(\boldsymbol{R''}, \boldsymbol{R'}) \\
            H^{EX}(\boldsymbol{R''}, \boldsymbol{R'})
        \end{pmatrix}
         = \int \phi_1^*(\boldsymbol{\rho_1}, \boldsymbol{\lambda_1}) \phi_2^*(\boldsymbol{\rho_2}, \boldsymbol{\lambda_2}) \delta(\boldsymbol{R} - \boldsymbol{R''})  \nonumber \\
       \times \begin{pmatrix}
            1\mathcal{A''} \\
            H\mathcal{A''}
        \end{pmatrix}
          \phi_1(\boldsymbol{\rho_1}, \boldsymbol{\lambda_1}) \phi_2(\boldsymbol{\rho_2}, \boldsymbol{\lambda_2}) \nonumber \\
         \delta(\boldsymbol{R} - \boldsymbol{R'})  \, d\boldsymbol{\rho_1} \, d\boldsymbol{\lambda_1} \, d\boldsymbol{\rho_2} \, d\boldsymbol{\lambda_2} \, d\boldsymbol{R}
\end{align}
	 So the Eq \eqref{phi6q} can be written as
	\begin{align}
		\int L(\boldsymbol{R''},\boldsymbol{R'})\chi(\boldsymbol{R'})\,d\boldsymbol{R'}=0
		\label{RGM}
	\end{align}
	where
	\begin{align}
		L(\boldsymbol{R''},\boldsymbol{R'})=&H(\boldsymbol{R''},\boldsymbol{R'})-EN(\boldsymbol{R''},\boldsymbol{R'}) \nonumber \\
		=&\left[-\frac{\nabla_{\boldsymbol{R'}}^2}{2\mu}+V_{rel}^D \left(\boldsymbol{R'}\right)-E_{rel}\right]\delta\left(\boldsymbol{R''}-\boldsymbol{R'}\right) \nonumber \\
		&+H^{EX}\left(\boldsymbol{R''}, \boldsymbol{R'}\right)-EN^{EX}\left(\boldsymbol{R''}, \boldsymbol{R'}\right)
	\end{align}	
	In the above equation, the reduced mass for the two-quark system is defined as $\mu$. $E_{rel}=E-E_{int}$ denotes the relative motion energy, and $V_{rel}^D$ represents the direct term in the interaction potential.
	
	The Eq \eqref{RGM} is a differential-integral equation which is difficult to solve. Therefore, We will expand the relative motion wave function $\chi_{L}(\boldsymbol{R})$ using a series of known Gaussian wave functions with different generating coordinates $\boldsymbol{S}_i$ ($i=1,~2,~\cdots n$):
	\begin{align}
		\chi_{L}(\boldsymbol{R}) &= \frac{1}{\sqrt{4\pi}} \left(\frac{3}{2\pi b^2}\right)^{3/4} \sum_{i=1}^n C_i  \nonumber \\
		&\quad \times \int \exp\left[-\frac{3}{4b^2} \left(\boldsymbol{R} - \boldsymbol{S}_i\right)^2 \right] Y_{LM}(\hat{\boldsymbol{S}}_i) \, d\hat{\boldsymbol{S}}_i  \nonumber \\
		=& \sum_{i=1}^n C_i \frac{u_{L}(\boldsymbol{R},\boldsymbol{S_{i}})}{R} Y_{LM}(\hat{\boldsymbol{R}})
		\label{chichichi}
	\end{align}
	with
	\begin{align}
	\frac{u_L (\boldsymbol{R},\boldsymbol{S_{i}})}{ R}=\sqrt{4\pi}(\frac{3}{2\pi b^2})^{\frac{3}{4}} e^{-\frac{3}{4 b^2}( \boldsymbol{R}^2+ \boldsymbol{r}_{i}^2)}i^L j_L(-i \frac{3}{2b^2}Rr_{i})
	\label{20}
	\end{align}
	where $C_i$ represent the expansion coefficients, $n$ denotes the number of the Gaussian bases, which is determined by the stability of the results, and $j_L$ represents the $L$th spherical Bessel function. Then
the relative motion wave function $\chi(\boldsymbol{R})$ is
   \begin{align}
   \chi(\boldsymbol{R}) &= \frac{1}{\sqrt{4\pi}} \sum_L \left(\frac{3}{2\pi b^2}\right)^{3/4} \sum_{i=1}^n C_{i,L}  \nonumber \\
		&\quad \times \int \exp\left[-\frac{3}{4b^2} \left(\boldsymbol{R} - \boldsymbol{S}_i\right)^2 \right] Y_{LM}(\hat{\boldsymbol{S}}_i) \, d\Omega_{{\boldsymbol{S}}_i}
   \end{align}
   Since the system we studied are all $S$-waves, $L=0$ in this work.
	The center of mass motion wave function is:
	\begin{align}
		Z(\boldsymbol{R}_c) = \left(\frac{6}{\pi b^2}\right)^{3/4} e^{-\frac{3\boldsymbol{R}_c^2}{b^2}}
		\label{zzzzz}
	\end{align}
	And for the orbital wave functions, $\phi_\alpha(\boldsymbol{S}_i)$ and $\phi_\beta(-\boldsymbol{S}_i)$ represent single-particle orbital wave functions centered at different reference points:
	\begin{align}
		\phi_\alpha(\boldsymbol{S}_i) &= \left(\frac{1}{\pi b^2}\right)^\frac{3}{4} e^{-\frac{(\boldsymbol{r_\alpha} - \boldsymbol{S}_i/2)^2}{2b^2}}, \nonumber \\
		\phi_\beta(-\boldsymbol{S}_i) &= \left(\frac{1}{\pi b^2}\right)^\frac{3}{4} e^{-\frac{(\boldsymbol{r_\beta} + \boldsymbol{S}_i/2)^2}{2b^2}}
		\label{phiphiphi}
	\end{align}
	
	The Eq \eqref{phi5q} can be rewritten by substituting Eq \eqref{chichichi}, \eqref{zzzzz}, and \eqref{phiphiphi}:
	\begin{align}
		\Psi_{6q} &=\sum_{i=1}^n C_i \psi_{i}\\
		&= \mathcal{A} \sum_{i=1}^n C_i \int \frac{d\hat{\boldsymbol{S}}_i}{\sqrt{4\pi}}
		\prod_{\alpha=1}^3 \phi_\alpha(\boldsymbol{S}_i) \prod_{\beta=4}^6 \phi_\beta(-\boldsymbol{S}_i)  \nonumber \\
		&\quad \times \left[ \left[\phi_{I_1 S_1}(B_1) \phi_{I_2 S_2}(B_2)\right]^{IS}
		Y_{LM}(\hat{\boldsymbol{S}}_i) \right]^J  \nonumber \\
		&\quad \times \left[\chi_c(B_1) \chi_c(B_2)\right]^{[\sigma]},
	\end{align}
The detail of constructing the wave functions ($\phi_{I_i S_i}$ and $\chi_c$) are presented in Appendix \ref{WAVE FUNCTION}.
	
	Finally, substituting Eq \eqref{chichichi} into Eq \eqref{phi6q} leads to the algebraic form of the RGM equation:
	\begin{align}
		\sum_j C_j H_{i,j} = E \sum_j C_j N_{i,j}.
    \label{EC}
	\end{align}
	where $H_{i,j}$ is the Hamiltonian matrix elements ($H_{ij}=\langle\psi_{i} | H |\psi_{j}\rangle$) and $N_{i,j}$ is the overlap ($N_{ij}=\langle\psi_{i} |\psi_{j}\rangle$). By solving Eq. \eqref{EC}, the energy and wave function of the six quark system can be obtained.
	
	For a scattering problem, the relative wave function is
expanded as:

 \begin{align}
 \chi_{L}(\boldsymbol{R})=& \sum_{i=1}^n c_i \frac{\tilde{u}_{L}(\boldsymbol{R},\boldsymbol{S}_{i})}{R} Y_{LM}(\hat{\boldsymbol{R}})
 \end{align}

 The wave function for a scattering state must satisfy the following boundary condition:
 \begin{align}
 \tilde{u}_{L}(\boldsymbol{R},\boldsymbol{S}_{i}) =
    \begin{cases}
        0, &  \boldsymbol{R} = 0 \\
        \left[ h_L^{(-)}( \boldsymbol{k}, \boldsymbol{R}) + S_L \, h_L^{(+)}( \boldsymbol{k}, \boldsymbol{R}) \right] {\boldsymbol{R}}, &  \boldsymbol{R} >  \boldsymbol{R}_C
    \end{cases}
     \label{u_L}
 \end{align}

 In this formulation, the functions $h_L^{(\pm)}$ are the Hankel functions. The symbol $k$ denotes the wave number of the relative motion, $k=\frac{\sqrt{2 \mu E_{cm}}}{\hbar}$ ($E_{cm}$ represents the kinetic energy of their relative motion). The parameter $ \boldsymbol{R}_C$ is the cutoff radius, for $ \boldsymbol{R} >  \boldsymbol{R}_C$, the strong interaction is considered negligible. The scattering matrix $S_L$ is related to the scattering phase shifts $\delta_{L}$:
 \begin{align}
 S_{L} = |S_{L}| \, e^{2i\delta_{L}}
 \label{sl}
 \end{align}
	To determine the scattering phase shifts, it is necessary to solve for the scattering matrix $S_{L}$ and the wave function $\tilde{u}_{L}(\boldsymbol{R},\boldsymbol{S}_{i})$. We begin by introducing a trial wave function $\tilde{u}_{t}(\boldsymbol{R},\boldsymbol{S}_{i})$ that satisfies Eq (\ref{u_L}), and then
	the trial wave function is expanded into a series of known basis functions $\tilde{u}_{i}(\boldsymbol{R},\boldsymbol{S}_{i})$
	\begin{align}
	\tilde{u}_t(\boldsymbol{R},\boldsymbol{S}_{i} ) = \sum_{i=0}^{n} c_i \tilde{u}_i( \boldsymbol{R}, \boldsymbol{S}_{i})
	\label{ut}
	\end{align}
	  This wave function $\tilde{u}_{i}(\boldsymbol{R},\boldsymbol{S}_{i})$ satisfies the condition:
	\begin{align}
	\tilde{u}_i( R, S_{i}) =
\begin{cases}
\alpha_i u_i^{(\text{in})}( \boldsymbol{R}, \boldsymbol{S}_{i}), &  \boldsymbol{R} \leq  \boldsymbol{R}_C \\
\left[h_L^{(-)}(\boldsymbol{k}, \boldsymbol{R}) + S_i h_L^{(+)}(\boldsymbol{k}, \boldsymbol{R})\right] \boldsymbol{R}, &  \boldsymbol{R} >  \boldsymbol{R}_C
\end{cases}
\label{uir}
	\end{align}
	where the function $u_i^{(\text{in})}( \boldsymbol{R}, \boldsymbol{S}_{i})$ is from Eq (\ref{20}).
	
	The Eq (\ref{ut}) can be rewritten by substituting Eq (\ref{uir}):
	\begin{align}
	&\tilde{u}_t( \boldsymbol{R}, \boldsymbol{S}_{i}) \nonumber \\
	& =
\begin{cases}
\sum c_i \alpha_i u_i^{(\text{in})}(\boldsymbol{ R}, \boldsymbol{S}_{i}), &  \boldsymbol{R} \leq \ \boldsymbol{R}_C \\
\sum  \left[c_ih_L^{(-)}(\boldsymbol{k}, \boldsymbol{R}) + c_iS_i h_L^{(+)}(\boldsymbol{k}, \boldsymbol{R})\right]  \boldsymbol{R}, &  \boldsymbol{R} > \boldsymbol{R}_C
\end{cases}
\label{uttt}
	\end{align}
By imposing continuity of both the wave function and its derivative at $ \boldsymbol{R}= \boldsymbol{R}_C$ in the formual (\ref{uir}), we obtain a system of two equations. This allows us to solve for the two unknown quantities $\alpha_i$ and $S_i$. The next step is to solve for the coefficients $c_i$. 	
	
	From the Eq (\ref{uttt}), we can obtain the follow equation:
	\begin{align}
	\sum_{i=0}^{n} c_i &= 1  \label{ci}\\
    \sum_{i=0}^{n} c_i S_i&= S_t
	\end{align}
	The Eq (\ref{ci}) can be rewritten as:
	\begin{align}
	c_0=1-\sum_{i=1}^{n} c_i
	\end{align}
	The Eq (\ref{ut}) can be written as:
	\begin{align}
	\tilde{u}_t( \boldsymbol{R}, \boldsymbol{S}_{i}) = \tilde{u}_0( \boldsymbol{R}, \boldsymbol{S}_{i}) + \sum_{i=1}^{n} c_i \left[ \tilde{u}_i( \boldsymbol{R}, \boldsymbol{S}_{i}) -\tilde{ u}_0( \boldsymbol{R}, \boldsymbol{S}_{i}) \right]
	\end{align}
	Substituting the above into the Schrödinger projection equation $\langle \delta \Psi' | H - E | \Psi \rangle = 0$, we can obtain
the formula:
\begin{align}
\sum_{j=1}^n \mathcal{L}_{ij} c_j = \mathcal{M}_i, \quad (i = 1, \dots, n)
\end{align}	
	where
	\begin{align}
	\mathcal{L}_{ij} &= \mathcal{K}_{ij} - \mathcal{K}_{i0} - \mathcal{K}_{0j} + \mathcal{K}_{00}
	\end{align}
	\begin{align}
\mathcal{M}_i &= \mathcal{K}_{00} - \mathcal{K}_{i0}
\end{align}
\begin{align}
\mathcal{K}_{ij} = \langle \phi_A(\xi_A) \phi_B(\xi_B) \tilde{u}_i( \boldsymbol{R}, \boldsymbol{S}_{i})/  \boldsymbol{R} \cdot Y_{LM}(\boldsymbol{\hat{R}}) \nonumber\\
| H - E | \mathcal{A}[\phi_A(\xi_A) \phi_B(\xi_B) \tilde{u}_j( \boldsymbol{R}, \boldsymbol{S}_{i})/  \boldsymbol{R} \cdot Y_{LM}(\boldsymbol{\hat{R}})] \rangle
	\end{align}
	By solving the above equations, we can obtain the expansion coefficients $c_i$, and subsequently calculate the scattering matrix elements $S_L$:
\begin{equation}
S_L = S_t - i \alpha \sum_{i=0}^n \mathcal{K}_{0i} c_i.
\end{equation}
 where $\alpha=\frac{\mu}{\hbar k}$.
	 Finally, we determine the scattering phase shifts $\delta_{L}$ through the Eq (\ref{sl}).
	
In the case of multi-channel coupling, the total wave function can be expressed as:
\begin{align}
\Psi^{(c)} = \sum_{\gamma} \mathcal{A}[\Phi_{\gamma} \chi_{\gamma}^{(c)}( \boldsymbol{R}_{\gamma})] + \Omega^{(c)} \quad (c = \alpha, \beta)
	\end{align}
	In this expression, the symbol $\gamma$ denotes all two-body channels, and $\Omega^{(c)}$ represents the decay residual amplitude not included in the previous term, which can generally be omitted. $\Psi^{(c)}$ describes the relative motion wave function for both the incident channel $c$ and all outgoing channels. The asymptotic behavior
of $\chi_{\gamma}^{(c)}( \boldsymbol{R}_{\gamma})$ is:
\begin{align*}
\chi_{\gamma}^{(c)}( \boldsymbol{R}_{\gamma}) &= \chi_{\gamma}^{(-)}( \boldsymbol{k}_{\gamma}, \boldsymbol{R}_{\gamma})\delta_{\gamma c} + {S}_{\gamma c}\chi_{\gamma}^{(+)}(\boldsymbol{k}_{\gamma}, \boldsymbol{R}_{\gamma}),~\boldsymbol{R}_{\gamma} > \boldsymbol{R}_{\gamma}^{(c)}
\end{align*}
	Here, ${S}_{\gamma c}$ denotes the $S$-matrix element corresponding to the transition $c \rightarrow \gamma$, $\chi_{\gamma}^{\pm}$ satisfies the following conditions:\\
	
	(1) For open channels:
	\begin{equation}
\chi_{\alpha}^{(\pm)}(\boldsymbol{k}_{\alpha}, \boldsymbol{R}_{\alpha}) = \frac{1}{\sqrt{v_{\alpha}}} h_{L_{\alpha}}^{(\pm)}(\boldsymbol{k}_{\alpha}, \boldsymbol{R}_{\alpha}), \quad \boldsymbol{R}_{\alpha} > \boldsymbol{R}_{\alpha}^{C}
\end{equation}
	
	(2) For closed channels:
	\begin{equation}
\chi_{\alpha}^{(\pm)}(\boldsymbol{k}_{\alpha}, \boldsymbol{R}_{\alpha}) = W_{L_{\alpha}}^{(\pm)}(\boldsymbol{k}_{\alpha}, \boldsymbol{R}_{\alpha}), \quad \boldsymbol{R}_{\alpha} > \boldsymbol{R}_{\alpha}^{C}
\end{equation}
	where $v_{\alpha}=\frac{\hbar k_{\alpha}}{\mu_{\alpha}}$ is the relative velocity, and the functions $W_{L_{\alpha}}^{(\pm)}(\boldsymbol{k}_{\alpha}, \boldsymbol{R}_{\alpha})$ are given in Ref. \cite{MKA}.
	
	The scattering matrix elements for the multi-channel systems can be computed using the similar method of the single-channel calculation.
	
	
	\section{The results and discussions}
	\label{11}

In this work, we systematically investigate $S-$wave singly bottomed ($b=1$) dibaryon states with the angular momentum $J=1$, isospin $I=0$ and strangeness $S=-1,~-3,~-5$. In the following discussion, we will first investigate the interaction between two baryons by computing the effective potential, and then carry out bound state calculations on the system to look for possible dibaryon states. Finally, we also investigate the scattering process to search for the resonance states.

\subsection{Effective potentials}

In particle physics, the effective potential is a simplified model for particle interactions. It is constructed by integrating out the microscopic degrees of freedom of the system. The process reduces the complex many body dynamics into a simplified effective potential, which describes the interactions between the remaining effective particles \cite{V}.

\begin{figure}[H]
		\centering
		\begin{subfigure}{1.0\linewidth}
			\includegraphics[width=\linewidth]{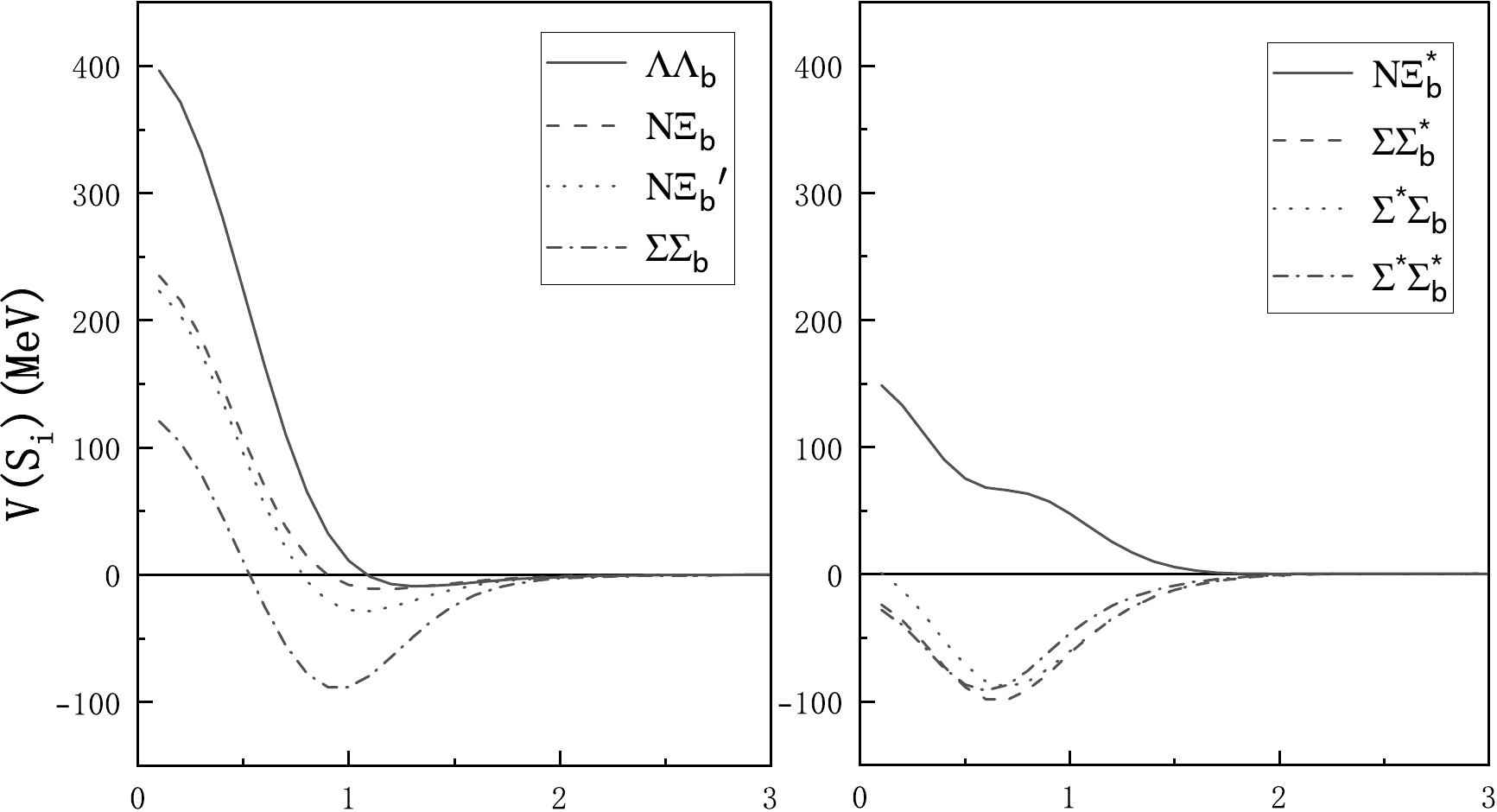}
		\end{subfigure}
		\[
		\mathrm{S_{i}\left(fm\right)} \quad \text{}
		\]
		\caption{The effective potentials of different channels for singly bottomed dibaryons with $S=-1$.}
		\label{Figure1}
	\end{figure}

\begin{figure}[H]
		\centering
		\begin{subfigure}{1\linewidth}
			\includegraphics[width=\linewidth]{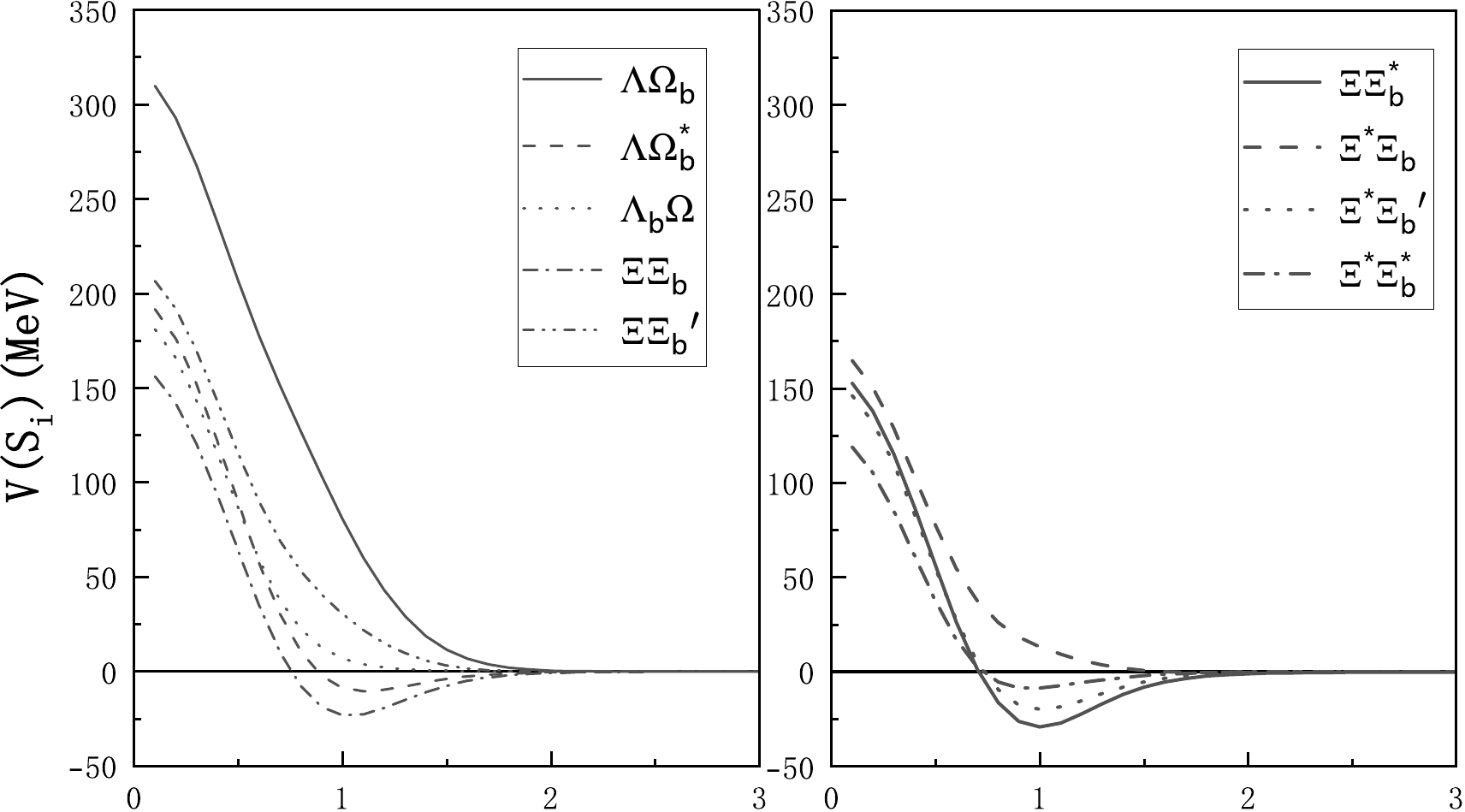}
		\end{subfigure}
		\[
		\mathrm{S_{i}\left(fm\right)} \quad \text{}
		\]
		\caption{The effective potentials of different channels for singly bottomed dibaryons with $S=-3$.}
		\label{Figure2}
	\end{figure}

\begin{figure}[H]
		\centering
		\begin{subfigure}{1.0\linewidth}
			\includegraphics[width=\linewidth]{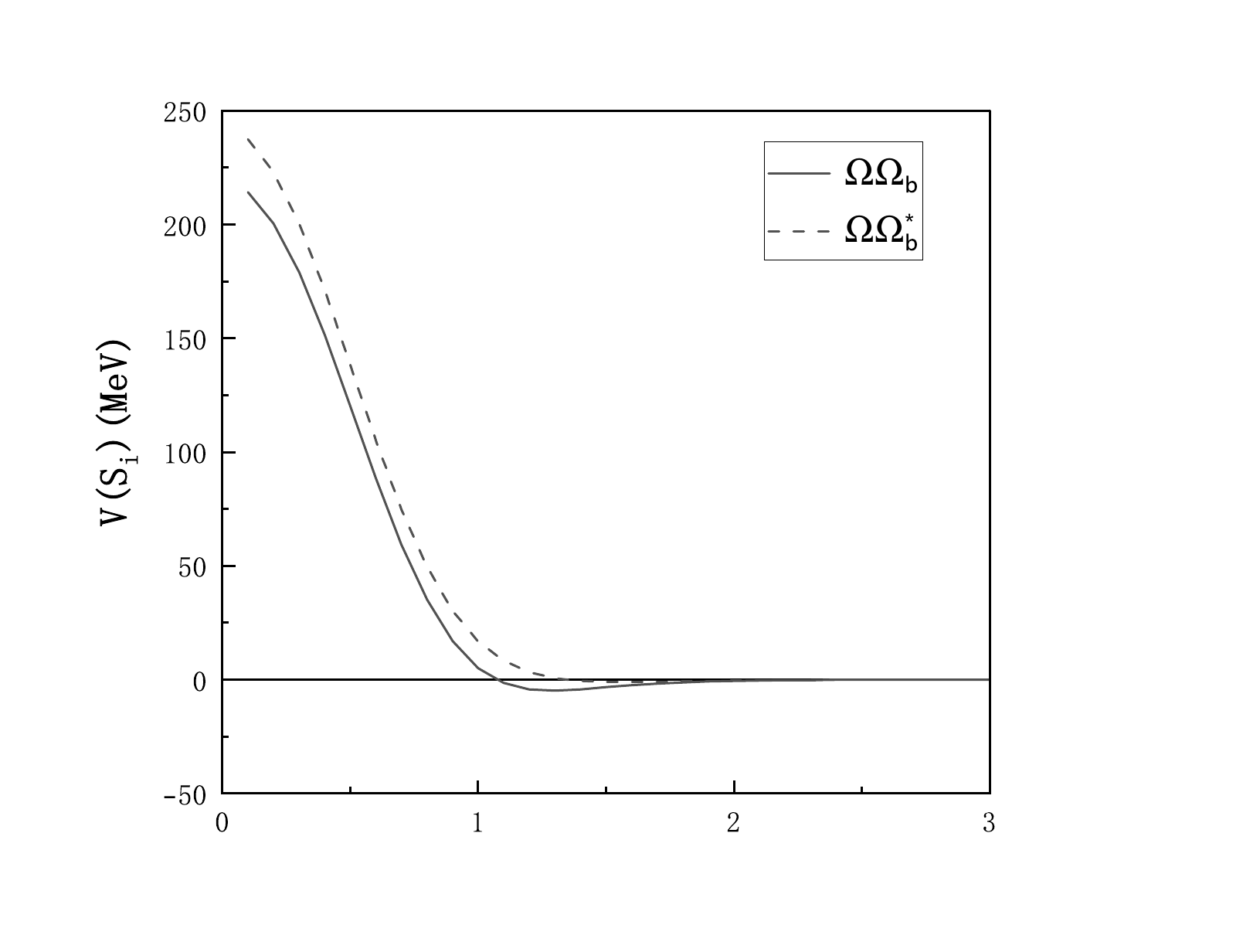}
		\end{subfigure}
		\caption{The effective potentials of different channels for singly bottomed dibaryons with $S=-5$.}
		\label{Figure3}
	\end{figure}

	In this work, the effective potential between the two baryons can be expressed in the following form:
	\begin{align}
		V\left(S_i\right)=E\left(S_i\right)-E\left(\infty\right)
	\end{align}
	The symbol $S_i$ denotes the distance between the two baryon clusters, $E\left(\infty\right)$ represents the interaction energy at a sufficiently large separation distance, and $E\left(S_i\right)$ is explicitly given by:
	\begin{equation}
		E(S_i) = \frac{\langle \Psi_{6q}(S_i) \, | \, H \, | \, \Psi_{6q}(S_i) \rangle}{\langle \Psi_{6q}(S_i) | \Psi_{6q}(S_i) \rangle}
	\end{equation}
	$\Psi_{6q}(S_i)$ represents the specific wave function of a dibaryon state, $\langle\Psi_{6q}(S_i)\,|\,H\,|\,\Psi_{6q}(S_i)\rangle$ and $\langle \Psi_{6q}(S_i) | \Psi_{6q}(S_i) \rangle$ are Hamiltonian matrix and the overlap of the states.~The effective potentials for systems with varying strange quarks are presented in Figs.~\ref{Figure1}-\ref{Figure3}. Although most of them have attractive and repulsive regions, as long as there is a negative minimum potential between two hadrons, it indicates the existence of an attractive potential between these two hadrons, which may form a bound state.

 For the $S=-1$ system, the effective potentials of different channels are shown in Fig.~\ref{Figure1}.~Among these eight channels, only the ${N\Xi^{*}_{b}}$ channel exhibits a purely repulsive interaction, while the other seven channels all display attractive potentials. Additionally, the effective potentials of the $\Sigma\Sigma_{b}$, $\Sigma\Sigma^{*}_{b}$, $\Sigma^{*}\Sigma_{b}$ and $\Sigma^{*}\Sigma^{*}_{b}$ are deeper than the other three channels, indicating that these four channels are more prone to form bound states. Especilally, the deepest attractive potentials of $\Sigma\Sigma^{*}_{b}$ channel is more than 100 MeV at a distance of around 0.6 fm between the two clusters, which means that it is more likely to form bound state.

 For the $S=-3$ system, the effective potentials of different channels are shown in Fig.~\ref{Figure2}. The $\Xi\Xi_{b}$, $\Lambda\Omega^{*}_{b}$, $\Xi\Xi^{*}_{b}$, $\Xi^{*}\Xi_{b}^{\prime}$, $\Xi^{*}\Xi_{b}^{*}$ channels show attractive interactions, while the other four channels have repulsive potentials. Also, the effective potentials of $\Xi\Xi_{b}$, $\Xi\Xi^{*}_{b}$ and $\Xi^{*}\Xi_{b}^{\prime}$ have deeper potentials well compared to $\Lambda\Omega^{*}_{b}$ and $\Xi^{*}\Xi_{b}^{*}$ channels, showing that these three channels are more prone to form bound states.

For the $S=-5$ system, the effective potentials of different channels are shown in Fig.~\ref{Figure3}.~We can identify two distinct channels in this system.~Both ${\Omega\Omega_{b}}$ and ${\Omega\Omega^{*}_{b}}$ states exist weakly attractive interactions. Consequently, we conclude that the two channels may be difficult to form bound states.

From the study of the effective potential above, it can be observed that some states exhibit deeply attractive interactions, some show weakly attractive interactions, while others are purely repulsive interactions. These findings provide valuable insights for the subsequent research on bound states and resonance states.

\subsection{Bound state calculation}
To investigate possible bound states, we conducted systematic bound state calculations. We calculated the energies of each single channel. Meanwhile, we also consider the effect of channel coupling and perform coupled channel calculations, and the computational results are listed in Tables \ref{tab1}-\ref{tab3}.
In these Tables, the first column indicates the particle species; the second column $E_{{th}}$ shows the corresponding theoretical thresholds; the third column $E_{sc}$ presents the single channel energies, while the fourth column $B_{sc}$ displays the single channel binding energies ($B_{sc}=E_{sc}-E_{{th}}$). The fifth $E_{cc}$ column gives the lowest energies of the system after channel coupling, and the final column $B_{cc}$ provides the binding energies from channel coupling,  where $B_{cc}=E_{cc}-E^{min}_{{th}}$ ($E^{min}_{{th}}$ represents the lowest theoretical thresholds). For states which no bound, we denote them with 'ub' (unbound).

For the \textbf{$S=-1$} system, the energies of different channels are shown in Table \ref{tab1}. Within this system, the single channel energy calculations demonstrate that four channels $\Sigma\Sigma_{b}$, $\Sigma\Sigma^{*}_{b}$, $\Sigma^{*}\Sigma_{b}$ and $\Sigma^{*}\Sigma^{*}_{b}$ can form bound states with the binding energies of -18.23 MeV, -30.51 MeV, -21.28 MeV, -24.05 MeV, respectively. The remaining four channels exhibit energies above corresponding the theoretical thresholds, which confirms that no bound states are formed in these channels. The results are consistent with the predictions from effective potential analysis. The effective potentials of the four channels ($\Sigma\Sigma_{b}$, $\Sigma\Sigma^{*}_{b}$, $\Sigma^{*}\Sigma_{b}$ and $\Sigma^{*}\Sigma^{*}_{b}$) in Fig.~\ref{Figure1} are approaching 100 MeV, which indicates that the attraction in these four channels is strong enough to form bound states. Although the effective potentials in the three channels ($\Lambda\Lambda_{b},~N\Xi_{b},~N\Xi_{b}^{\prime}$) are attractive, the attraction is too weak to form bound states. The remaining channel ($N\Xi_{b}^{*}$) exhibits a purely repulsive effective potential, which prevents the bound state formation. Coupled channel energy calculations further reveal that the system ground state energy remains higher than the minimum theoretical threshold $\left(\Lambda\Lambda_{b}\right)$. We therefore conclude that no bound states below the lowest threshold exist in this system. However, there may exist resonance states. For the higher energy single channel bound states ($\Sigma\Sigma_{b}$, $\Sigma\Sigma^{*}_{b}$, $\Sigma^{*}\Sigma_{b}$ and $\Sigma^{*}\Sigma^{*}_{b}$), they may decay into the corresponding open channels through coupling with the open channels $\left(\Lambda\Lambda_{b},~N\Xi_{b},~N\Xi_{b}^{\prime},~N\Xi_{b}^{*}\right)$, and the existence as resonance states may be verified through scattering processes, which will be discussed in section \ref{Resonance states}. \\

\begin{table}[H]
		\centering
		\renewcommand{\arraystretch}{1.5}
		\caption{The energy (in MeV) of different channels for singly bottomed dibaryons with $S=-1$.}
		\label{tab1}
		\begin{tabular}{lccccc}
			\hline
			\hline
			Channels & $E_{\text{th}}$  & $E_{\text{sc}}$  & $B_{\text{sc}}$  & $E_{\text{cc}}$  & $B_{\text{cc}}$ \\
		\hline
			$\Lambda \Lambda_b$    & 6713.23    &~~6719.50   &~~ub    &    &    \\
			\cline{1-4}
			$N\Xi_b$    & 6713.15   &~~6719.11   &~~ub    &    &    \\
			\cline{1-4}
			$N\Xi'_b$    & 6835.67   &~~6839.70   &~~ub    &    &    \\
			\cline{1-4}
			$N\Xi^*_b$    & 6848.79   &~~6856.78    &~~ub    &~~ \multirow{2}{*}{6718.48} &   ~~\multirow{2}{*}{ub} \\
			\cline{1-4}
			$\Sigma\Sigma_b$    & 6996.29    &~~6978.06 &~~-18.23    &~~    &~~    \\
			\cline{1-4}
			$\Sigma \Sigma^*_b$    & 7011.48    &~~6980.97 &~~-30.51  &    &     \\
			\cline{1-4}
			$\Sigma^* \Sigma_b$    & 7193.19    &~~7171.91 &~~-21.28 &    &    \\
			\cline{1-4}
			$\Sigma^* \Sigma^*_b$    & 7208.38   &~~7184.33 &~~-24.05 &    &    \\
			\hline
			\hline
		\end{tabular}
	\end{table}

For the $S=-3$ system, the energies of different channels are shown in Table \ref{tab2}. We observe that no bound states are formed in these channels. All single channel calculation results lie above their respective theoretical thresholds. After performing coupled channel calculation, the system ground state energy 7020.23 MeV remains above the minimum threshold 7017.84 MeV ($\Xi\Xi_b$), again demonstrating no bound states exist. The results are physically reasonable as clearly shown in Fig.~\ref{Figure2}. The effective attraction in the five channels ($\Xi\Xi_{b}$, $\Lambda\Omega^{*}_{b}$, $\Xi\Xi^{*}_{b}$, $\Xi^{*}\Xi_{b}^{\prime}$, $\Xi^{*}\Xi_{b}^{*}$ ) is insufficient to form any bound states, while the remaining channels exhibit purely repulsive effective potentials.\\

\begin{table}[H]
	\centering
	\renewcommand{\arraystretch}{1.5}
	\caption{The energy (in MeV) of different channels for singly bottomed dibaryons with $S=-3$.}
	\label{tab2}
	\begin{tabular}{lccccc}
		\hline
		\hline
		Channels & $E_{\text{th}}$ & $E_{\text{sc}}$ & $B_{\text{sc}}$ & $E_{\text{cc}}$  & $B_{\text{cc}}$  \\
		\hline
		$\Lambda \Omega_b$    & 7020.21    &~~7027.05    &~~ub    &    &    \\
		\cline{1-4}
		$\Lambda \Omega^*_b$    & 7031.26    &~~7036.37   &~~ub    &    &    \\
		\cline{1-4}
		$\Lambda_b\Omega$    & 7209.29    &~~7214.71    &~~ub    &    &    \\
		\cline{1-4}
		$\Xi\Xi_b$    & 7017.84 &~~7021.45   &~~ub   &~~ \multirow{2}{*}{7020.23} &   ~~\multirow{2}{*}{ub}     \\
		\cline{1-4}
		$\Xi\Xi'_b$    & 7140.35   &~~7146.47    &~~ub   \\
		\cline{1-4}
		$\Xi\Xi^*_b$    & 7153.48   &~~7156.03    &~~ub    &   &     \\
		\cline{1-4}
		$\Xi^* \Xi_b$    & 7214.74    &~~7220.42    &~~ub    &    &    \\
		\cline{1-4}
		$\Xi^* \Xi'_b$    & 7337.25   &~~7341.34    &~~ub    &    &    \\
		\cline{1-4}
		$\Xi^* \Xi^*_b$    & 7350.38   &~~7355.22   &~~ub    &    &    \\
		\hline
		\hline
	\end{tabular}
\end{table}

For the \textbf{$S=-5$} system, the energies of different channels are shown in Table \ref{tab3}. We observe that all single channel energies remain above their respective theoretical thresholds. Coupled channel calculation further demonstrates that the system ground state energy of 7520.10 MeV exceeds the lowest 7516.26 MeV threshold ($\Omega \Omega_b$), confirming the absence of bound states even with channel coupling effect. The results are physically reasonable as clearly demonstrated in Fig.~\ref{Figure3}. The effective attraction in these two channels ($\Omega \Omega_b$, $\Omega \Omega^*_b$) is insufficient to form any bound states.

\begin{table}[H]
	\centering
	\renewcommand{\arraystretch}{1.5}
	\caption{The energy (in MeV) of different channels for singly bottomed dibaryons with $S=-5$.}
	\label{tab3}
	\begin{tabular}{lccccc}
		\hline
		\hline
		Channels & $E_{\text{th}}$  & $E_{\text{sc}}$  & $B_{\text{sc}}$  & $E_{\text{cc}}$ & $B_{\text{cc}}$  \\
		\hline
		$\Omega \Omega_b$    & 7516.26   &~~7520.10    &~~ub    &~~   \multirow{2}{*}{7520.10} &   ~~\multirow{2}{*}{ub}     \\
		\cline{1-4}
		$\Omega \Omega^*_b$    & 7527.32    &~~7531.47  &~~ub    &~~  \\
		\hline
		\hline
	\end{tabular}
\end{table}

  \subsection{Resonance states}
  \label{Resonance states}
  As mentioned above, due to the strong attractive forces in the system, some channels are bound. These states can decay into the corresponding open channels through coupling with them, forming resonance states.~To further investigate the existence of resonance states, we examine the scattering phase shifts of all possible open channels.~This work considers two types of channel coupling.~One involves the two-channel coupling with an open state and a bound channel, the other is the five-channel system comprising an open channel and four bound states. This section focuses on the presence of the single channel bound state with $S=-1$, while no bound states are found for the $S=-3$ and $S=-5$. We will discuss the possible existence of resonance states by examining the scattering phase shifts in various open channels. 

   The scattering phase shifts typically exhibit a smooth, continuous curve that varies gradually with increasing incident energy. In contrast, when a resonance state occurs, the phase shifts undergo an abrupt jump of approximately 180 degrees at a specific energy. Specifically, without considering the mass shift, the resonance energy $M^{\prime}$ is defined as the sum of incident energy, which the phase shifts reach 90 degrees, and the open channel threshold. Furthermore, the decay width $\Gamma_{i}$ is estimated by calculating the difference in incident energies corresponding to phase shifts of 45 and 135 degrees. $\Gamma_{total}$ is the total decay width of the resonance state.

\begin{figure}[H]
		\centering
		\begin{subfigure}[b]{0.02\textwidth} 
        \rotatebox{90}{\scriptsize{Phase Shifts(deg.)}} 
        \vspace{2cm} 
    \end{subfigure}%
		\begin{subfigure}{0.9\linewidth}
			\includegraphics[width=\linewidth]{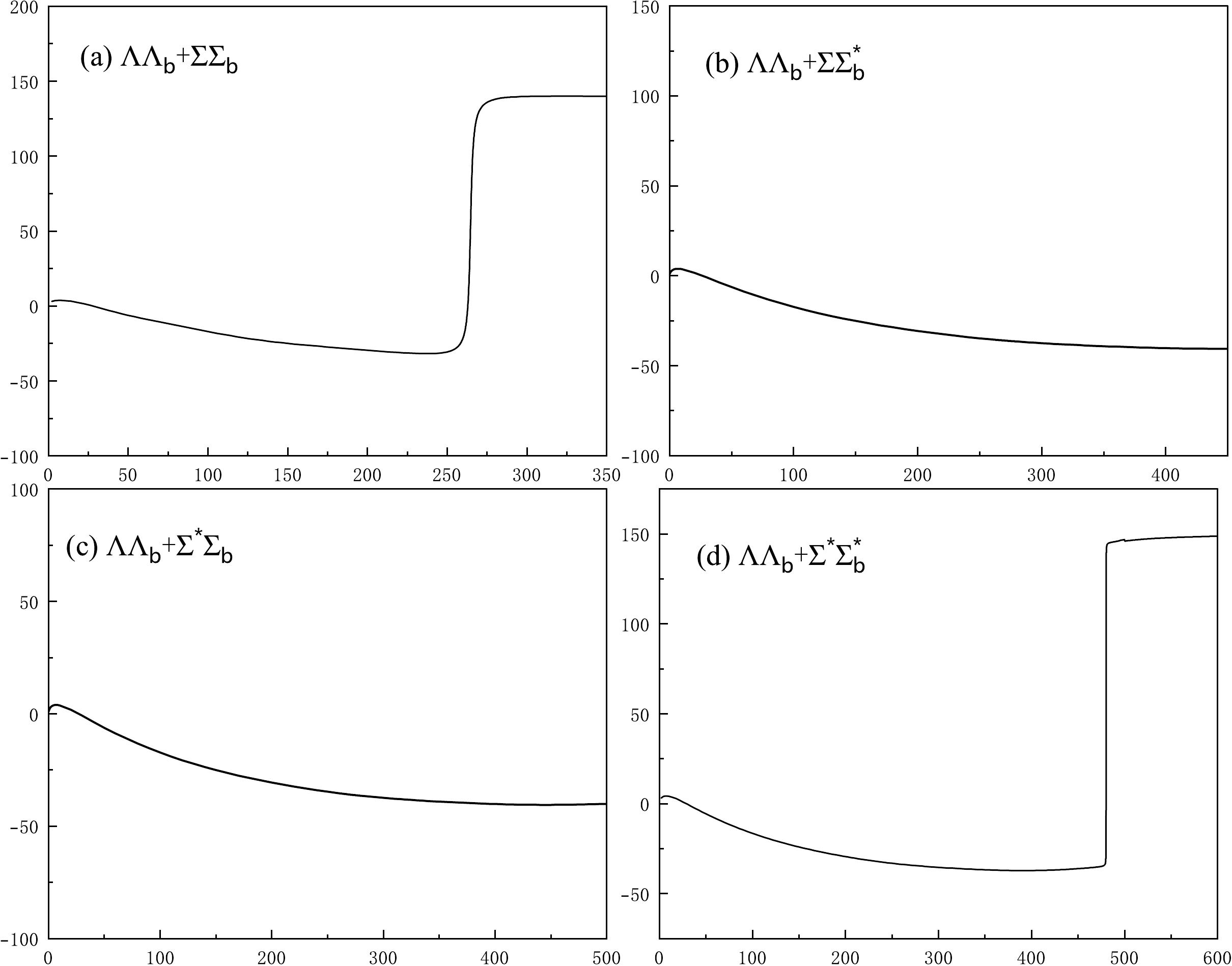}
		\end{subfigure}
		\[
		\mathrm{E_{c.m.}\left(MeV\right)} \quad \text{}
		\]
		\caption{The $\Lambda \Lambda_b$ phase shifts with two-channel coupling.}
		\label{Figure4}
	\end{figure}

\begin{figure}[H]
		\centering
		\begin{subfigure}[b]{0.02\textwidth} 
        \rotatebox{90}{\scriptsize{Phase Shifts(deg.)}} 
        \vspace{2cm} 
    \end{subfigure}%
		\begin{subfigure}{0.9\linewidth}
			\includegraphics[width=\linewidth]{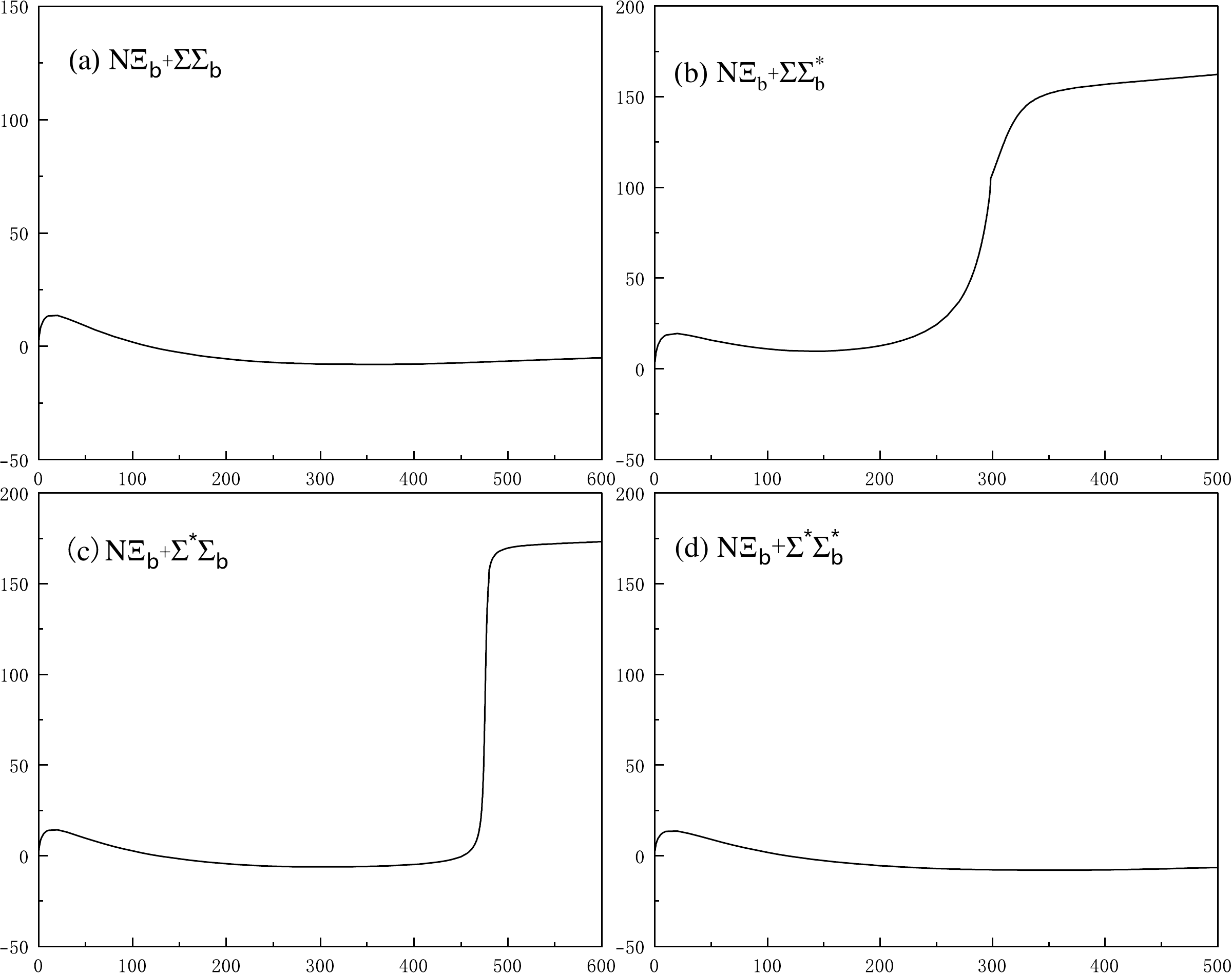}
		\end{subfigure}
		\[
		\mathrm{E_{c.m.}\left(MeV\right)} \quad \text{}
		\]
		\caption{The $N\Xi_b$ phase shifts with two-channel coupling.}
		\label{Figure5}
	\end{figure}

\begin{figure}[H]
		\centering
		\begin{subfigure}[b]{0.02\textwidth} 
        \rotatebox{90}{\scriptsize{Phase Shifts(deg.)}} 
        \vspace{2cm} 
    \end{subfigure}%
		\begin{subfigure}{0.9\linewidth}
			\includegraphics[width=\linewidth]{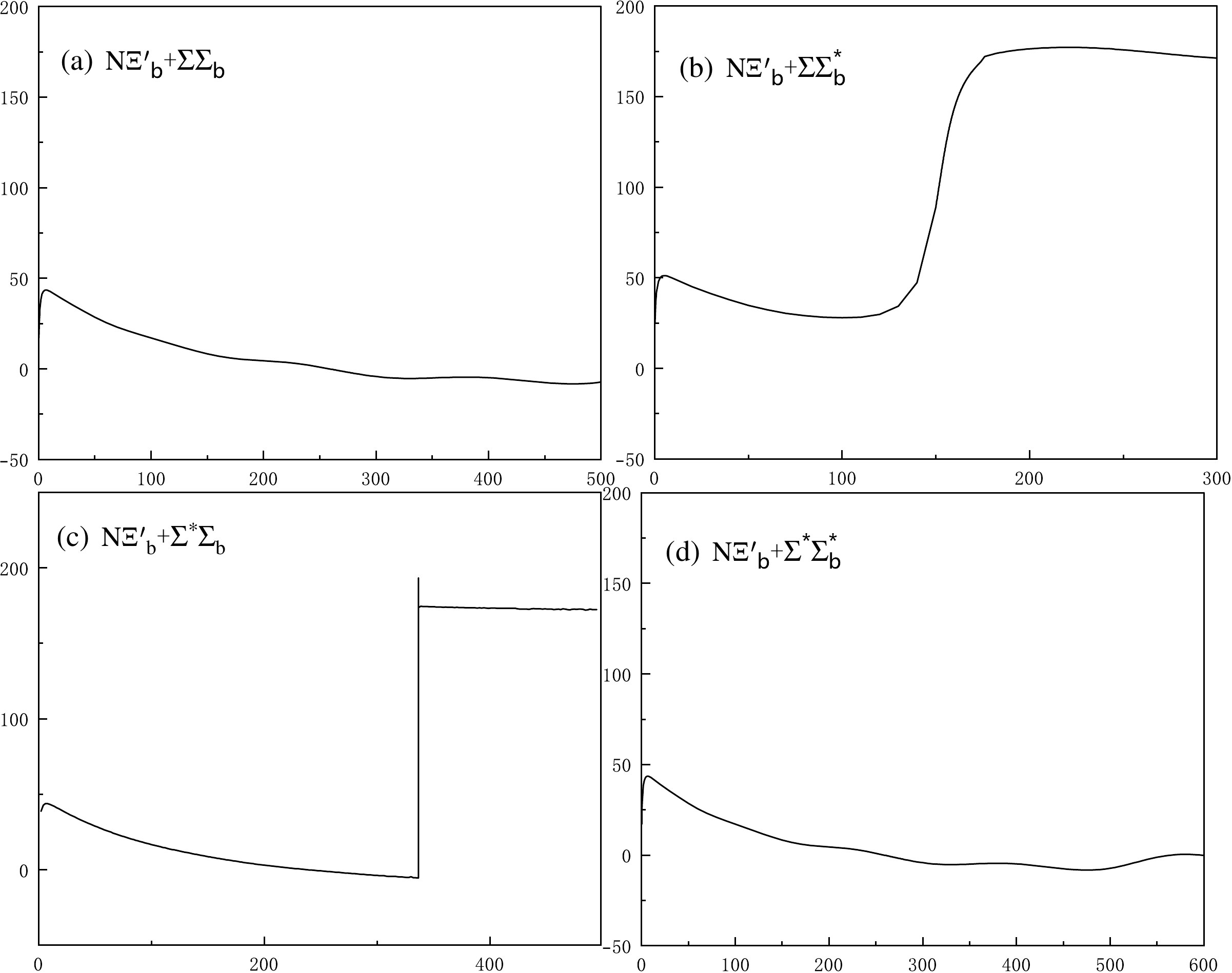}
		\end{subfigure}
		\[
		\mathrm{E_{c.m.}\left(MeV\right)} \quad \text{}
		\]
		\caption{The $N\Xi'_b$ phase shifts with two-channel coupling.}
		\label{Figure6}
	\end{figure}

\begin{figure}[H]
		\centering
		\begin{subfigure}[b]{0.02\textwidth} 
        \rotatebox{90}{\scriptsize{Phase Shifts(deg.)}} 
        \vspace{2cm} 
    \end{subfigure}%
		\begin{subfigure}{0.9\linewidth}
			\includegraphics[width=\linewidth]{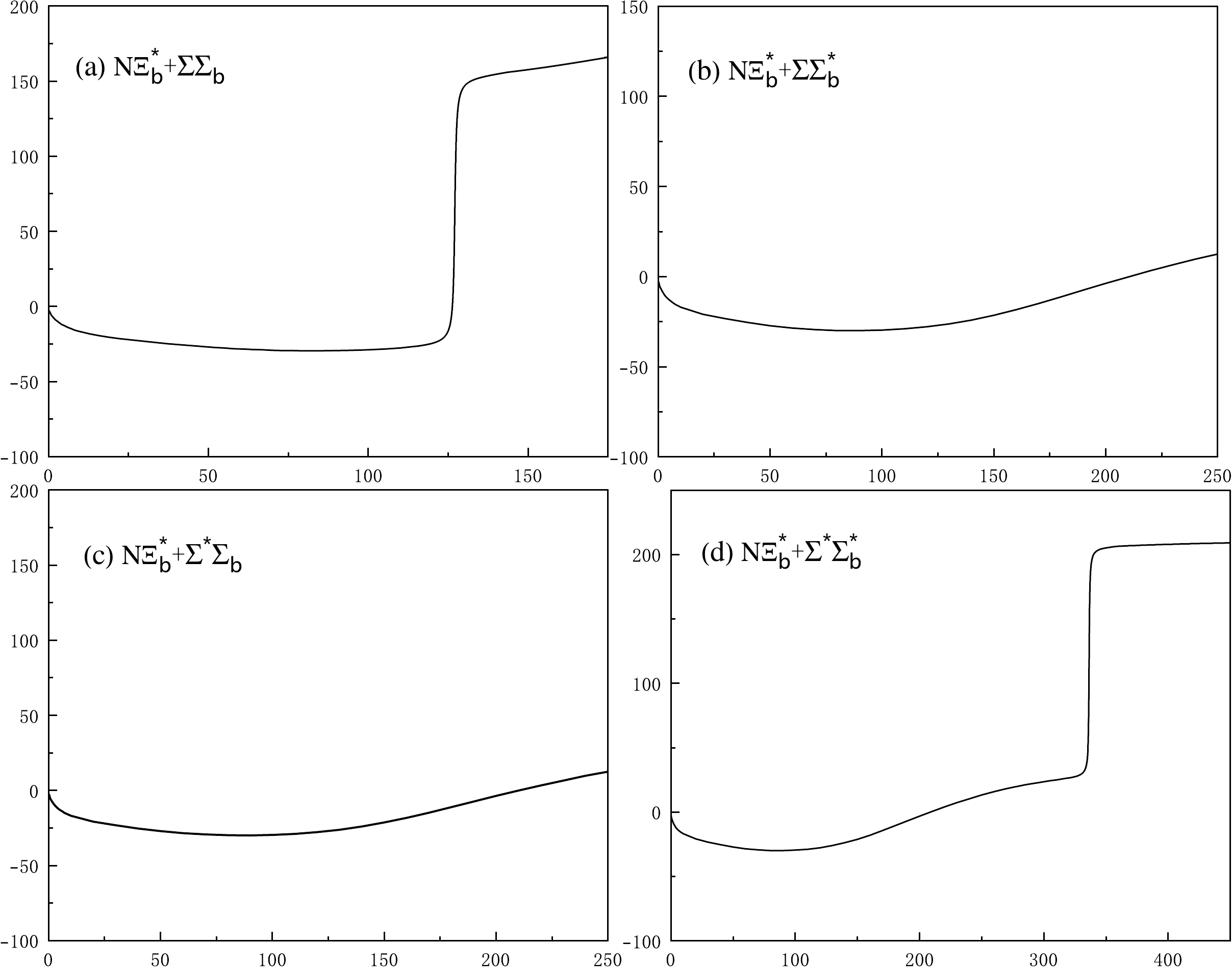}
		\end{subfigure}
		\[
		\mathrm{E_{c.m.}\left(MeV\right)} \quad \text{}
		\]
		\caption{The $N\Xi_b^{*}$ phase shifts with two-channel coupling.}
		\label{Figure7}
	\end{figure}

\begin{figure}[H]
		\centering
		\begin{subfigure}[b]{0.02\textwidth} 
        \rotatebox{90}{\scriptsize{Phase Shifts(deg.)}} 
        \vspace{2cm} 
    \end{subfigure}%
		\begin{subfigure}{0.9\linewidth}
			\includegraphics[width=\linewidth]{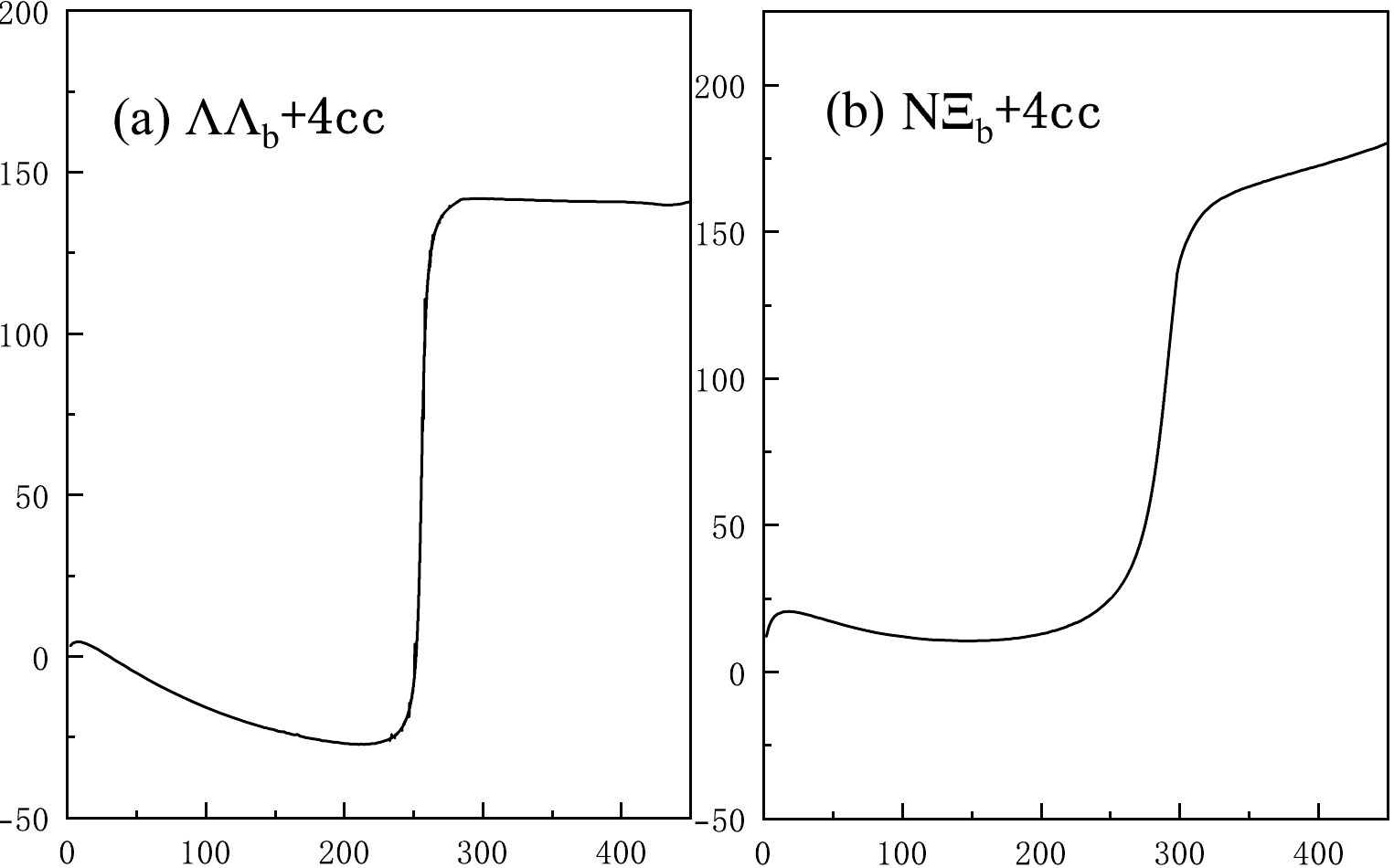}
		\end{subfigure}
		\[
		\mathrm{E_{c.m.}\left(MeV\right)} \quad \text{}
		\]
		\caption{The $\Lambda\Lambda_b$ and $N\Xi_b$ phase shifts with five-channel coupling.}
		\label{Figure8}
	\end{figure}

\begin{figure}[H]
		\centering
		\begin{subfigure}[b]{0.02\textwidth} 
        \rotatebox{90}{\scriptsize{Phase Shifts(deg.)}} 
        \vspace{2cm} 
    \end{subfigure}%
		\begin{subfigure}{0.9\linewidth}
			\includegraphics[width=\linewidth]{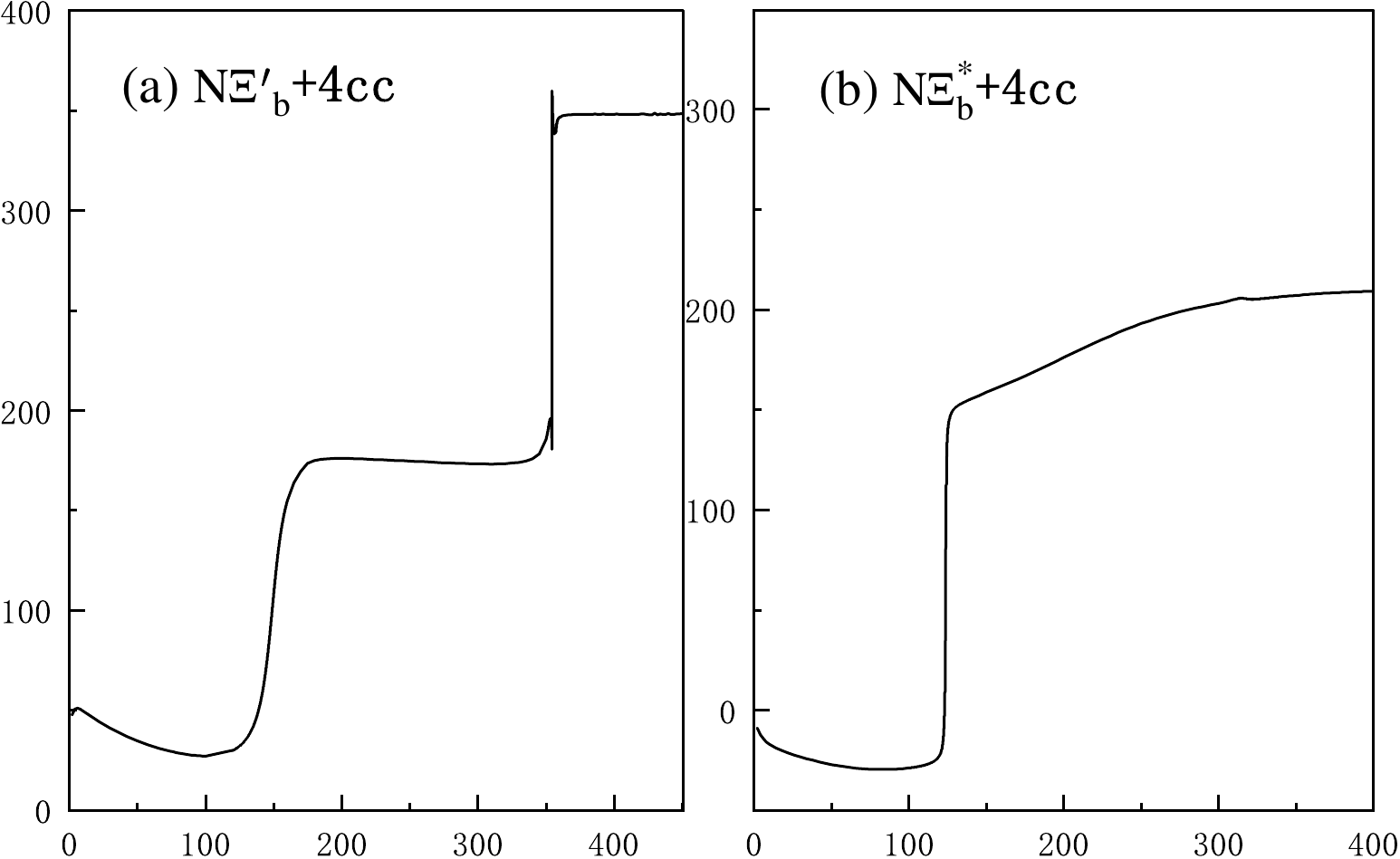}
		\end{subfigure}
		\[
		\mathrm{E_{c.m.}\left(MeV\right)} \quad \text{}
		\]
		\caption{The $N\Xi'_b$ and $N\Xi_b^*$ phase shifts with five-channel coupling.}
		\label{Figure9}
	\end{figure}

  The scattering phase shifts of the two-channel coupling are presented in Figs.~\ref{Figure4}-\ref{Figure7}, and those of the five-channel coupling are presented in Fig.~\ref{Figure8} and Fig.~\ref{Figure9}.
  It should be noted that in these Figures, the horizontal axis $E_{c.m.}$ represents the incident energy, while the vertical axis shows the scattering phase shifts. The theoretical resonance mass $M^{\prime}$ is obtained by adding $E_{c.m.}$ to the theoretical threshold of the corresponding open channel. To minimize systematic errors and enable comparison with future experimental results, the experimental resonance energy $M$ can be calculated by employing this formula $M=M^{\prime}-E_{th}+E_{exp}$, where $E_{th}$ and $E_{exp}$ are the theoretical and experimental thresholds of the resonance state, respectively. Meanwhile, the resonance energies and decay widths of two-channel coupling and five-channel coupling are listed in Table \ref{tab:resonance} and Table \ref{five-channel coupling}, repectively. In the column of resonance energy, the values in parentheses are theoretical resonance energies $M^{\prime}$, and the values outside the parentheses are experimental resonance energies $M$.

   The scattering phase shifts of $\Lambda\Lambda_b$ coupling with corresponding closed channels are shown in Fig.~\ref{Figure4}. The decay widths and resonance energies are listed in Table \ref{tab:resonance}. The Fig.~\ref{Figure4}(a) displays the $\Lambda\Lambda_b$ phase shifts coupling with the $\Sigma\Sigma_b$ state. The phase shifts exhibit a sharp rise at an incident energy of approximately 265 MeV, indicating that $\Sigma\Sigma_b$ is a resonance state with the resonance energy 6982.54(6978.83) MeV and the decay width 9.400 MeV. Similarly, the phase shifts of $\Lambda\Lambda_b$ coupling with $\Sigma^*\Sigma^*_b$ are shown in Fig.~\ref{Figure4}(d), where a sharp rise at 480 MeV suggests that $\Sigma^* \Sigma^*_b$ is a resonance state with the resonance energy 7129.94(7193.32) MeV and the decay width 0.135 MeV. In contrast, Fig.~\ref{Figure4}(b) and Fig.~\ref{Figure4}(c) show no abrupt change in the $\Lambda \Lambda_b$ phase shifts by coupling with the $\Sigma\Sigma^*_b$ / $\Sigma^* \Sigma_b$. Therefore, neither $\Sigma\Sigma^*_b$ nor $\Sigma^* \Sigma_b$ behaves like a resonance state in the $\Lambda\Lambda_b$ phase shifts. Two possible factors contribute to this phenomenon. First, the strong coupling between the two channels pushes the energy of the bound state above the threshold, converting it into a scattering state. Second, the coupling is too weak for a resonance state to manifest. To discern the underlying cause, we compute the cross matrix elements between the $\Lambda\Lambda_b$ channel and the $\Sigma \Sigma^*_b$ / $\Sigma^* \Sigma_b$ channels. The calculated elements are found to be approaching zero, indicating that the above phenomenon is due to an excessively weak coupling between the channels.

   The phase shifts of $N\Xi_b$ coupling with corresponding closed channels are shown in Fig.~\ref{Figure5}. As shown in Fig.~\ref{Figure5}(b) and Fig.~\ref{Figure5}(c), the $N\Xi_b$ scattering phase shifts coupling with $\Sigma\Sigma^*_b$ / $\Sigma^* \Sigma_b$ exhibit a sharp change at incident energies of approximately 296 MeV and 495 MeV, which means that $\Sigma\Sigma^*_b$ and $\Sigma^* \Sigma_b$ are resonance states with the resonance energies of 7017.07(7009.55) MeV and 7121.92(7189.11) MeV, the decay widths of 39.700 MeV and 3.950 MeV. Conversely, no abrupt change occurs in the $N\Xi_b$ phase shifts by coupling with $\Sigma\Sigma_b$ / $\Sigma^* \Sigma^*_b$. 
   The cross matrix elements between $N\Xi_b$ and $\Sigma\Sigma_b$ / $\Sigma^* \Sigma^*_b$ are close to zero, which shows that the absence of resonance states is due to the weak channel coupling between $N\Xi_b$ and $\Sigma\Sigma_b$ / $\Sigma^*\Sigma^*_b$, which is insufficient to produce resonance states.
	
   The phase shifts of $N\Xi'_b$ state coupling with corresponding closed channels are shown in Fig.~\ref{Figure6}. The decay widths and resonance energies are listed in Table \ref{tab:resonance}. From the Fig.~\ref{Figure6}(b), it can be seen that the scattering phase shifts of $N\Xi'_b$ coupling with $\Sigma\Sigma^*_b$ change abruptly at an incident energy of approximately 150 MeV, concluding that the $\Sigma\Sigma^*_b$ is a resonance state with the resonance energy 6993.39(6985.87) MeV and decay width 17.700 MeV. Meanwhile, the phase shifts of $N\Xi'_b$ coupling with $\Sigma^*\Sigma_b$ displayed in Fig.~\ref{Figure6}(c) show a very sharp rise at 336 MeV, suggesting that the $\Sigma^*\Sigma_b$ is a resonance state with the resonance energy 7105.03(7172.22) MeV and a very narrow decay width 0.001 MeV.
     Conversely, no abrupt change occurs in the $N\Xi'_b$ phase shifts by coupling with $\Sigma\Sigma_b$ / $\Sigma^*\Sigma^*_b$ channels.
     The cross matrix elements between $N\Xi'_b$ channels and $\Sigma\Sigma_b$ / $\Sigma^*\Sigma^*_b$ are computed. These elements are found to be approaching zero, which confirms that an excessively weak coupling between the channels is responsible for the phenomenon.

	The scattering phase shifts of $N\Xi_b^{*}$ coupling with corresponding closed channels are shown in Fig.~\ref{Figure7}. The decay widths and resonance energies are listed in Table \ref{tab:resonance}. As shown in Fig.~\ref{Figure7}(a), the $N\Xi_b^{*}$ phase shifts coupling with $\Sigma\Sigma_b$ exhibit a sharp change at an incident energy of approximately 127 MeV, demonstrating that $\Sigma\Sigma_b$ channel is a resonance state with the resonance energy 6979.77(6976.06) MeV and the decay width 1.290 MeV. Additionally, the phase shifts of $N\Xi_b^{*}$ coupling with $\Sigma^*\Sigma_b^{*}$ displayed in Fig.~\ref{Figure7}(d) show an abrupt change at an incident energy of 336 MeV, which means that $\Sigma^*\Sigma^*_b$ is a resonance state with the resonance energy 7121.71(7185.09) MeV and decay width 1.300 MeV.
	In contrast, the Fig.~\ref{Figure7}(b) and Fig.~\ref{Figure7}(c) show no abrupt change in the $N\Xi_b^{*}$ phase shifts by coupling with $\Sigma\Sigma^*_b$ / $\Sigma^*\Sigma_b$ channels. 
	The cross matrix elements between $N\Xi_b^{*}$ and $\Sigma\Sigma^*_b$ / $\Sigma^* \Sigma_b$ are computed and found to be nearly zero. This confirms that the effect arises from exceptionally weak coupling.
	
	To account for the influence between different closed channels on resonance states, all closed channels are coupled. Consequently, the scattering phase shifts in a single open channel are analyzed under the inclusion of four closed channels.
  The five-channel phase shifts are presented in Fig.~\ref{Figure8} and Fig.~\ref{Figure9}, and the decay widths and resonance energies are listed in Table \ref{five-channel coupling}.

  The five-channel coupling phase shifts of $\Lambda\Lambda_b$ are shown in the Fig.~\ref{Figure8}(a), only one abrupt change is observed at an incident energy of approximately 257 MeV. The corresponding resonance energy is 6974.22(6970.51) MeV, which shows that this resonance state corresponds to $\Sigma\Sigma_b$ and the decay width is 13.410 MeV. It can be seen that the resonance energy 6974.22(6970.51) MeV for the five-channel coupling is lower than that the resonance energy 6982.54(6978.83) MeV for the two-channel case. This occurs because coupling among closed channels can further lower the energies of low-lying channels, while pushing higher channels to even higher energies, sometimes even above the threshold. In contrast, the inclusion of open channels tends to increase the energy of the resonance state. Consequently, the coupling between open and multiple closed channels introduces a competitive effect on the resonance energy, which may either increase or decrease it.~Hence, the resonance state $\Sigma^*\Sigma^*_b$, which appears in the two-channel coupling, becomes a scattering state when coupled to the five-channel. 
	
	The five-channel coupling phase shifts of $N\Xi_b$ are shown in the Fig.~\ref{Figure8}(b), only one abrupt change is observed at an incident energy of approximately 287 MeV. The corresponding resonance energy is 7008.37(7000.85) MeV, suggesting that this resonance state corresponds to $\Sigma\Sigma_b^*$ with the decay width 26.070 MeV. Meanwhile, the coupling among closed channels further lowers the energies of low-lying channels while pushing higher channels to even higher energies. Therefore, the $\Sigma^*\Sigma_b$ state, which is presented in the two-channel coupling, transforms to the scattering state through the five-channel coupling. 

	\begin{table*}
\centering
\caption{The masses and decay widths of resonance states (in MeV) for different scattering processes for the two-channel coupling.}
\label{tab:resonance}
\begin{tabular}{@{}c*{16}{c}@{}}
\hline
\hline
 & \multicolumn{8}{c}{Two channel coupling}  \\
 \hline
\cmidrule(lr){2-9} \cmidrule(lr){10-17}
\multirow{2}{*}{Open channels} & \multicolumn{2}{c}{$\Sigma \Sigma_b$} & \multicolumn{2}{c}{$\Sigma \Sigma_b^*$} & \multicolumn{2}{c}{$\Sigma^* \Sigma_b$} & \multicolumn{2}{c}{$\Sigma^* \Sigma^*_b$} \\
 & M & $\Gamma_i$ & M & $\Gamma_i$ & M & $\Gamma_i$ & M & $\Gamma_i$ &\\
\hline
\hline
$\Lambda \Lambda_b$ & 6982.54(6978.83) &~~9.400 & \multicolumn{2}{c}{...} & \multicolumn{2}{c}{...} &7129.94(7193.32) &~~0.135  \\
\hline
$N\Xi_b$ & \multicolumn{2}{c}{...} &7017.07(7009.55) &~~39.700 &~~7121.92(7189.11)  &~~3.950 & \multicolumn{2}{c}{...}  \\
\hline
$N\Xi'_b$ & \multicolumn{2}{c}{...} & 6993.39(6985.87) &~~17.700 &~~7105.03(7172.22) &~~0.001 & \multicolumn{2}{c}{...}  \\
\hline
$N\Xi^*_b$ & 6979.77(6976.06) &~~1.290 & \multicolumn{2}{c}{...} & \multicolumn{2}{c}{...} & 7121.71(7185.09) &~~1.300 \\
\hline
$\Gamma_{total}$ & &~~10.690 & &~~57.400 & &~~3.951 & &~~1.435 \\
\hline
\hline
\end{tabular}
\end{table*}
\begin{table*}
\centering
\caption{The masses and decay widths of resonances (in MeV) for different scattering processes for the five-channel coupling.}
\label{five-channel coupling}
\begin{tabular}{@{}c*{16}{c}@{}}
\hline
\hline
 & \multicolumn{8}{c}{Five channel coupling}  \\
 \hline
\cmidrule(lr){2-9} \cmidrule(lr){10-17}
\multirow{2}{*}{Open channels} & \multicolumn{2}{c}{$\Sigma \Sigma_b$} & \multicolumn{2}{c}{$\Sigma \Sigma_b^*$} & \multicolumn{2}{c}{$\Sigma^* \Sigma_b$} & \multicolumn{2}{c}{$\Sigma^* \Sigma^*_b$} \\
 & M & $\Gamma_i$ & M & $\Gamma_i$ & M & $\Gamma_i$ & M & $\Gamma_i$ &\\
\hline
\hline
$\Lambda \Lambda_b$ & 6974.22(6970.51) &~~13.410 & \multicolumn{2}{c}{...} & \multicolumn{2}{c}{...}  & \multicolumn{2}{c}{...}  \\
\hline
$N\Xi_b$ & \multicolumn{2}{c}{...} & 7008.37(7000.85) &~~26.070  & \multicolumn{2}{c}{...} & \multicolumn{2}{c}{...}  \\
\hline
$N\Xi'_b$ & \multicolumn{2}{c}{...} & 6990.69(6983.17) &~~17.720 &~~7122.94(7190.13) &~~0.010 & ~~~~~&~~~~~~ \\
\hline
$N\Xi^*_b$ & 6975.37(6971.66) & 1.040 & \multicolumn{2}{c}{...} & \multicolumn{2}{c}{...} & \multicolumn{2}{c}{...}  \\
\hline
$\Gamma_{total}$ & &~~14.450 & &~~43.790 & & & & \\
\hline
\hline
\end{tabular}
\end{table*}
The five-channel coupling phase shifts of $N\Xi'_b$ are exhibited in the Fig.~\ref{Figure9}(a). Unlike the previous two cases, there are two abrupt changes in the scattering phase shifts. The first one occurs at an incident energy of 147 MeV, corresponding to the resonance state $\Sigma \Sigma_b^*$ with the resonance energy of 6990.69(6983.17) MeV. The second one has a very abrupt change occuring at an incident energy of 354 MeV, corresponding to the resonance state $\Sigma^*\Sigma_b$ with the resonance energy 7122.94(7190.13) MeV. And their decay widths are 17.720 MeV and 0.010 MeV, respectively.

The five-channel coupling phase shifts of $N\Xi_b^*$ are presented in the Fig.~\ref{Figure9}(b). Only one abrupt change is observed at an incident energy of approximately 122 MeV. The corresponding resonance energy is 6975.37(6971.66) MeV, demonstrating that this resonance state corresponds to $\Sigma\Sigma_b$ with the decay width 1.040 MeV. The coupling among closed channels further lowers the energies of low-lying channels, while pushing higher channels to even higher energies. Therefore, the resonance state $\Sigma^*\Sigma^*_b$, which is presented in the two-channel coupling, evolves into the scattering state in the five-channel coupling. 

Based on the above discussion, we obtain two resonance states finally. The one is $\Sigma\Sigma_b$ with the experimental resonance energy 6974.22 MeV - 6975.37 MeV and the decay width is 14.450 MeV. The other one is $\Sigma \Sigma_b^*$ with the experimental resonance energy 6990.69 MeV - 7008.37 MeV and the decay width is 43.790 MeV.
Although the $\Sigma^*\Sigma_b$ behaves like a resonance state in the $N\Xi'_b$ scattering process, it can transform into scattering state through the $N\Xi_b$ channel. Therefore, $\Sigma^*\Sigma_b$ cannot be regarded as a resonance state. 
Additionlly, the $\Sigma^*\Sigma^*_b$ behaves as a resonance state in the two-channel coupling system, but it disappears in the five-channel coupling system, which indicates that the effect of the channel-coupling cannot be ignored in the description of multiquark systems. 

Ultimately, to explore the structure of the resonance states, we also calculate the Root Mean Square (RMS) radius of the resonance states obtained. The result shows that the $\Sigma\Sigma_b$ state has a RMS radius of value 0.99 fm, which corresponds to a compact hexaquark configuration, distinguishing it from the deuteron-like structure. In contrast, the $\Sigma \Sigma_b^*$ state exhibits a RMS radius of value 1.77 fm, consistents with a molecular-like state analogous to the deuteron. Hence, $\Sigma \Sigma_b^*$ can be appropriately described as a deuteron-like dibaryon.


	\section{Summary}
	\label{22}
	
	In this work, we systematically investigate the existence of the deuteron-like singly bottomed dibaryon resonance states with strangeness $S=-1,~-3,~-5$ in the chiral quark model.
Construction of the dibaryon wave functions serves as the starting point. Subsequently, the interactions between two baryons are analyzed through the calculation of effective potential, which provide initial insights into the possibility of bound state. To verify the existence of bound states, we conduct calculations of the energy in both single and coupled channels. Additionlly, to search for resonance states, we investigate the scattering processes of the open channel coupling with closed channels. Finally, the Root Mean Square (RMS) radius of the resonance states are calculated to explore the structure of the resonance states. 
	
	 The numerical results show there is no any bound state or resonance state in the system with with strangeness $S=-3,~-5$. In the system with $S=-1$, the attractive effective potentials for the $\Sigma\Sigma_{b}$, $\Sigma\Sigma^{*}_{b}$, $\Sigma^{*}\Sigma_{b}$ and $\Sigma^{*}\Sigma^{*}_{b}$ states are deep enough to form bound states. However, when coupled to open channels, these states transform into resonance states or scattering states. From the study of the scattering phases of the open channels including the effects of channel-coupling, we obtain two resonance states. The first candidate is $\Sigma\Sigma_b$, which is a compact hexaquark with the resonance energy 6974.22 MeV - 6975.37 MeV and the decay width 14.450 MeV, respectively, and the possible decay channles are $\Lambda\Lambda_b$ and $N\Xi_b^*$. The other one is $\Sigma \Sigma_b^*$, which is a deuteron-like dibaryon resonance with the resonance energy 6990.69 MeV - 7008.37 MeV and the decay width 43.790 MeV, respectively, and the possible decay channles are $N\Xi_b$ and $N\Xi'_b$ channels. We also found that some singly bound states become scattering states by considering the channel-couping calcuation in the scattering process, which indicates that the effect of the channel-coupling cannot be neglected in the description of multiquark systems. Therefore, we emphasize that the study of scattering process and the coupling effect between open and closed channels is of great importance in investigating exotic hadron states.
	
	This work presents a comprehensive investigation of singly bottomed dibaryon states, more dibaryons containing heavy quarks deserve the systematic study. Besides, this work focuses on studying dibaryons from the perspective of hadron-hadron structure. Some studies have started from the diquark structure to investigate the dibaryons containing heavy quarks, and have obtained some possible bound states\cite{db,sh}. Therefore, studying various types of dibaryons of different structures will also be a continuing research task for us in the future.
	We also hope the theoretical study will receive experimental confirmation, to stimulate the progress in dibaryon research and to deep the theoretical understanding of QCD.

\acknowledgments{This work is supported partly by the National Natural Science Foundation of China under Contracts Nos. 12575088, 11675080, 11775118 and 11535005.}	
	
	\appendix
	\section{\label{Gell-Mann matrices}The Gell-Mann matrices}
	
	For the $SU(5)$ group describing flavor degrees of freedom, the number of generators $T^a=\frac{1}{2}\lambda^{a}$ is 24, and the specific forms $\lambda^{a}$ are:
		
		\begin{align*}
\lambda^{1} &= \begin{pmatrix}
  0 & 1 & 0 & 0 & 0 \\
  1 & 0 & 0 & 0 & 0 \\
  0 & 0 & 0 & 0 & 0 \\
  0 & 0 & 0 & 0 & 0 \\
  0 & 0 & 0 & 0 & 0
\end{pmatrix},
\lambda^{2} = \begin{pmatrix}
  0 & -i & 0 & 0 & 0 \\
  i & 0 & 0 & 0 & 0 \\
  0 & 0 & 0 & 0 & 0 \\
  0 & 0 & 0 & 0 & 0 \\
  0 & 0 & 0 & 0 & 0
\end{pmatrix},\\
\lambda^{3} &= \begin{pmatrix}
  1 & 0 & 0 & 0 & 0 \\
  0 & -1 & 0 & 0 & 0 \\
  0 & 0 & 0 & 0 & 0 \\
  0 & 0 & 0 & 0 & 0 \\
  0 & 0 & 0 & 0 & 0
\end{pmatrix},
\lambda^{4} = \begin{pmatrix}
  0 & 0 & 1 & 0 & 0 \\
  0 & 0 & 0 & 0 & 0 \\
  1 & 0 & 0 & 0 & 0 \\
  0 & 0 & 0 & 0 & 0 \\
  0 & 0 & 0 & 0 & 0
\end{pmatrix}, \\
\lambda^{5} &= \begin{pmatrix}
  0 & 0 & -i & 0 & 0 \\
  0 & 0 & 0 & 0 & 0 \\
  i & 0 & 0 & 0 & 0 \\
  0 & 0 & 0 & 0 & 0 \\
  0 & 0 & 0 & 0 & 0
\end{pmatrix},
\lambda^{6} = \begin{pmatrix}
  0 & 0 & 0 & 0 & 0 \\
  0 & 0 & 1 & 0 & 0 \\
  0 & 1 & 0 & 0 & 0 \\
  0 & 0 & 0 & 0 & 0 \\
  0 & 0 & 0 & 0 & 0
\end{pmatrix},\\
\lambda^{7} &= \begin{pmatrix}
  0 & 0 & 0 & 0 & 0 \\
  0 & 0 & -i & 0 & 0 \\
  0 & i & 0 & 0 & 0 \\
  0 & 0 & 0 & 0 & 0 \\
  0 & 0 & 0 & 0 & 0
\end{pmatrix},
\lambda^{8} = \frac{1}{\sqrt{3}}\begin{pmatrix}
  1 & 0 & 0 & 0 & 0 \\
  0 & 1 & 0 & 0 & 0 \\
  0 & 0 & -2 & 0 & 0 \\
  0 & 0 & 0 & 0 & 0 \\
  0 & 0 & 0 & 0 & 0
\end{pmatrix}, \\
\lambda^{9} &= \begin{pmatrix}
  0 & 0 & 0 & 1 & 0 \\
  0 & 0 & 0 & 0 & 0 \\
  0 & 0 & 0 & 0 & 0 \\
  1 & 0 & 0 & 0 & 0 \\
  0 & 0 & 0 & 0 & 0
\end{pmatrix},
\lambda^{10} = \begin{pmatrix}
  0 & 0 & 0 & -i & 0 \\
  0 & 0 & 0 & 0 & 0 \\
  0 & 0 & 0 & 0 & 0 \\
  i & 0 & 0 & 0 & 0 \\
  0 & 0 & 0 & 0 & 0
\end{pmatrix},\\
\lambda^{11} &= \begin{pmatrix}
  0 & 0 & 0 & 0 & 0 \\
  0 & 0 & 0 & 1 & 0 \\
  0 & 0 & 0 & 0 & 0 \\
  0 & 1 & 0 & 0 & 0 \\
  0 & 0 & 0 & 0 & 0
\end{pmatrix},
\lambda^{12} = \begin{pmatrix}
  0 & 0 & 0 & 0 & 0 \\
  0 & 0 & 0 & -i & 0 \\
  0 & 0 & 0 & 0 & 0 \\
  0 & i & 0 & 0 & 0 \\
  0 & 0 & 0 & 0 & 0
\end{pmatrix}, \\
\lambda^{13} &= \begin{pmatrix}
  0 & 0 & 0 & 0 & 0 \\
  0 & 0 & 0 & 0 & 0 \\
  0 & 0 & 0 & 1 & 0 \\
  0 & 0 & 1 & 0 & 0 \\
  0 & 0 & 0 & 0 & 0
\end{pmatrix},
\lambda^{14} = \begin{pmatrix}
  0 & 0 & 0 & 0 & 0 \\
  0 & 0 & 0 & 0 & 0 \\
  0 & 0 & 0 & -i & 0 \\
  0 & 0 & i & 0 & 0 \\
  0 & 0 & 0 & 0 & 0
\end{pmatrix},\\
\lambda^{15} &= \frac{1}{\sqrt{6}}\begin{pmatrix}
  1 & 0 & 0 & 0 & 0 \\
  0 & 1 & 0 & 0 & 0 \\
  0 & 0 & 1 & 0 & 0 \\
  0 & 0 & 0 & -3 & 0 \\
  0 & 0 & 0 & 0 & 0
\end{pmatrix},
\lambda^{16} = \begin{pmatrix}
  0 & 0 & 0 & 0 & 1 \\
  0 & 0 & 0 & 0 & 0 \\
  0 & 0 & 0 & 0 & 0 \\
  0 & 0 & 0 & 0 & 0 \\
  1 & 0 & 0 & 0 & 0
\end{pmatrix}, \\
\lambda^{17} &= \begin{pmatrix}
  0 & 0 & 0 & 0 & -i \\
  0 & 0 & 0 & 0 & 0 \\
  0 & 0 & 0 & 0 & 0 \\
  0 & 0 & 0 & 0 & 0 \\
  i & 0 & 0 & 0 & 0
\end{pmatrix},
\lambda^{18} = \begin{pmatrix}
  0 & 0 & 0 & 0 & 0 \\
  0 & 0 & 0 & 0 & 1 \\
  0 & 0 & 0 & 0 & 0 \\
  0 & 0 & 0 & 0 & 0 \\
  0 & 1 & 0 & 0 & 0
\end{pmatrix},\\
\end{align*}
\begin{align*}
\lambda^{19} &= \begin{pmatrix}
  0 & 0 & 0 & 0 & 0 \\
  0 & 0 & 0 & 0 & -i \\
  0 & 0 & 0 & 0 & 0 \\
  0 & 0 & 0 & 0 & 0 \\
  0 & i & 0 & 0 & 0
\end{pmatrix},
\lambda^{20} = \begin{pmatrix}
  0 & 0 & 0 & 0 & 0 \\
  0 & 0 & 0 & 0 & 0 \\
  0 & 0 & 0 & 0 & 1 \\
  0 & 0 & 0 & 0 & 0 \\
  0 & 0 & 1 & 0 & 0
\end{pmatrix}, \\
\lambda^{21} &= \begin{pmatrix}
  0 & 0 & 0 & 0 & 0 \\
  0 & 0 & 0 & 0 & 0 \\
  0 & 0 & 0 & 0 & -i \\
  0 & 0 & 0 & 0 & 0 \\
  0 & 0 & -i & 0 & 0
\end{pmatrix},
\lambda^{22} = \begin{pmatrix}
  0 & 0 & 0 & 0 & 0 \\
  0 & 0 & 0 & 0 & 0 \\
  0 & 0 & 0 & 0 & 0 \\
  0 & 0 & 0 & 0 & 1 \\
  0 & 0 & 0 & 1 & 0
\end{pmatrix},\\
\lambda^{23} &= \begin{pmatrix}
  0 & 0 & 0 & 0 & 0 \\
  0 & 0 & 0 & 0 & 0 \\
  0 & 0 & 0 & 0 & 0 \\
  0 & 0 & 0 & 0 & -i \\
  0 & 0 & 0 & i & 0
\end{pmatrix},
\lambda^{24} = \frac{1}{\sqrt{10}}\begin{pmatrix}
  1 & 0 & 0 & 0 & 0 \\
  0 & 1 & 0 & 0 & 0 \\
  0 & 0 & 1 & 0 & 0 \\
  0 & 0 & 0 & 1 & 0 \\
  0 & 0 & 0 & 0 & -4
\end{pmatrix}.\\
\end{align*}

		\section{\label{WAVE FUNCTION}The total wave function of dibaryon system}
	
	The total wave function of the dibaryon system is constructed by coupling the total wave functions of two baryons via the Clebsch-Gordan coefficients. In order to construct the wave function of the dibaryon, we must first construct the baryon wave functions.

For the baryon spin wave functions, the system involves a total of 8 distinct spin wave functions. The spin wave functions used are expressed:
\begin{align*}
	\chi_{\frac{3}{2}, \frac{3}{2}}^{\sigma} & = \alpha \alpha \alpha \nonumber \\
	\chi_{\frac{3}{2}, \frac{1}{2}}^{\sigma} & = \frac{1}{\sqrt{3}}(\alpha \alpha \beta+\alpha \beta \alpha+\beta \alpha \alpha) \nonumber \\
	\chi_{\frac{3}{2},-\frac{1}{2}}^{\sigma} & = \frac{1}{\sqrt{3}}(\alpha \beta \beta+\beta \alpha \beta+\beta \beta \alpha) \nonumber \\
    \chi_{\frac{3}{2},-\frac{3}{2}}^{\sigma} & = \beta \beta \beta \nonumber \\
    \chi_{\frac{1}{2},\frac{1}{2}}^{\sigma1} & = \sqrt{\frac{1}{6}}(2 \alpha \alpha \beta-\alpha \beta \alpha-\beta \alpha \alpha)  \nonumber\\
    \chi_{\frac{1}{2},\frac{1}{2}}^{\sigma2} & = \sqrt{\frac{1}{2}}(\alpha \beta \alpha-\beta \alpha \alpha)\nonumber\\
    \chi_{\frac{1}{2},-\frac{1}{2}}^{\sigma1} & = \sqrt{\frac{1}{6}}(\alpha \beta \beta+\beta \alpha \beta-2 \beta \beta \alpha) \nonumber \\
    \chi_{\frac{1}{2},-\frac{1}{2}}^{\sigma2} & = \sqrt{\frac{1}{2}}(\alpha \beta \beta-\beta \alpha \beta) \nonumber
\end{align*}
	We use $\chi_{S,S_{z}}^{\sigma}$ to denote the baryon spin wave function, where $S$ and $S_z$ represent the spin quantum number and its third component, respectively. For wave functions with identical quantum numbers but different symmetries, we distinguish them using different superscripts. For example: $\chi_{\frac{1}{2},\frac{1}{2}}^{\sigma 1}$ and $\chi_{\frac{1}{2},\frac{1}{2}}^{\sigma 2}$ represent symmetric and antisymmetric wave functions with spin quantum number $\frac{1}{2}$.
	
For baryon flavor wave functions: The system involves
a total of 46 distinct flavor wave function configurations. The flavor wave functions are displayed as follow:
\begin{align*}
			\chi_{0,0}^{f1} &= \frac{1}{2}(usd+sud-sdu-dsu) \\ 
			   \chi_{0,0}^{f2} &= \sqrt{\frac{1}{12}}(2uds-2dsu+sdu+usd-sud-dsu) \\
			   \chi_{0,0}^{f3}  & = \frac{1}{2}(ubd+bud-bdu-dbu)\\  
			    \chi_{0,0}^{f4} &= \sqrt{\frac{1}{12}}(2ubd-2dbu+bdu+ubd-bud-dbu)    \\
			     \chi_{0,0}^{f5}&= \sqrt{\frac{1}{6}}(2ssb-sbs-bss) \\  
			    \chi_{0,0}^{f6}&= \sqrt{\frac{1}{2}}(sbs-bss) \\
			    \chi_{0,0}^{f7} &= \sqrt{\frac{1}{3}}(ssb+sbs+bss) \\
			   \chi_{0,0}^{f8}  & = sss \\
			  \chi_{\frac{1}{2},-\frac{1}{2}}^{f1}  &=\frac{1}{2}(dbs+bds-bsd-sbd) \\
			   \chi_{\frac{1}{2},-\frac{1}{2}}^{f2}  &=\sqrt{\frac{1}{12}}(2dsb-2sdb+bsd+dbs-bds-sbd) \\
			    \chi_{\frac{1}{2},-\frac{1}{2}}^{f3}  & =\sqrt{\frac{1}{6}}(udd+dud-2ddu) \\
			      \chi_{\frac{1}{2},-\frac{1}{2}}^{f4}  & =\sqrt{\frac{1}{2}}(udd-dud) \\			   			
			       \chi_{\frac{1}{2},-\frac{1}{2}}^{f5}& =\sqrt{\frac{1}{12}}(2dsb+2sdb-bsd-dbs-bds-sbd) \\
			        \chi_{\frac{1}{2},-\frac{1}{2}}^{f6} & =\frac{1}{2}(dbs+sbd-bsd-bds) \\
			         \chi_{\frac{1}{2},-\frac{1}{2}}^{f7} & =\sqrt{\frac{1}{6}}(dss+sds-2ssd) \\
			          \chi_{\frac{1}{2},-\frac{1}{2}}^{f8}  & =\sqrt{\frac{1}{2}}(dss-sds) \\ 			
			          \chi_{\frac{1}{2},-\frac{1}{2}}^{f9}& =\sqrt{\frac{1}{6}}(dsb+sdb+bsd+dbs+bds+sbd) \\
			            \chi_{\frac{1}{2},-\frac{1}{2}}^{f10} & =\sqrt{\frac{1}{3}}(dss+sds+ssd) \\ 
			\chi_{\frac{1}{2},\frac{1}{2}}^{f1}  &=\sqrt{\frac{1}{6}}(2uud-udu-duu) \\
			              \chi_{\frac{1}{2},\frac{1}{2}}^{f2}&=\sqrt{\frac{1}{2}}(udu-duu) \\			
			              \chi_{\frac{1}{2},\frac{1}{2}}^{f3} & =\frac{1}{2}(ubs+bus-bsu-sbu)\\
			              \chi_{\frac{1}{2},\frac{1}{2}}^{f4} &=\sqrt{\frac{1}{12}}(2usb-2sub+bsu+ubs-bus-sbu)\\			
			              \chi_{\frac{1}{2},\frac{1}{2}}^{f5}  &=\sqrt{\frac{1}{12}}(2usb+2sub-bsu-ubs-bus-sbu)\\
			              \chi_{\frac{1}{2},\frac{1}{2}}^{f6}  &=\frac{1}{2}(ubs+sbu-bsu-bus) \\		
			              			     \end{align*}
			      \begin{align*}	
			             \chi_{\frac{1}{2},\frac{1}{2}}^{f7}  &=\sqrt{\frac{1}{6}}(uss+sus-2ssu)\\
			              \chi_{\frac{1}{2},\frac{1}{2}}^{f8} &=\sqrt{\frac{1}{2}}(uss-sus)\\			
			              \chi_{\frac{1}{2},\frac{1}{2}}^{f9}  &=\sqrt{\frac{1}{6}}(usb+sub+bsu+ubs+bus+sbu)\\
			             \chi_{\frac{1}{2},\frac{1}{2}}^{f10}  &=\sqrt{\frac{1}{3}}(uss+sus+ssu)\\
			              \chi_{1,-1}^{f1}   &= \sqrt{\frac{1}{6}}(2ddb-dbd-bdd)\\ 
			              \chi_{1,-1}^{f2}  &= \sqrt{\frac{1}{2}}(dbd-bdd)~~\\			
			              \chi_{1,-1}^{f3}  &= \sqrt{\frac{1}{6}}(2dds-dsd-sdd)~~\\ 
			              \chi_{1,-1}^{f4} &= \sqrt{\frac{1}{2}}(dsd-sdd)~~\\			
			              \chi_{1,-1}^{f5}  &= \sqrt{\frac{1}{3}}(ddb+dbd+bdd)~~\\
			              \chi_{1,-1}^{f6}  &= \sqrt{\frac{1}{3}}(dds+dsd+sdd)  \\
	\chi_{1,0}^{f1} &= \sqrt{\frac{1}{12}}(2uds+2dus-sdu-usd-sud-dsu)~~\\
			              \chi_{1,0}^{f2} &= \frac{1}{2}(usd+dsu-sdu-sud)~~\\			
			             \chi_{1,0}^{f3} &= \sqrt{\frac{1}{12}}(2udb+2dub-dbu-ubd-bud-dbu)~~\\
			              \chi_{1,0}^{f4}&= \frac{1}{2}(ubd+dbu-bdu-bud)~~\\			
			              \chi_{1,0}^{f5}  &= \sqrt{\frac{1}{6}}(udb+dub+bdu+ubd+bud+dbu)~~\\
			              \chi_{1,0}^{f6} &= \sqrt{\frac{1}{6}}(uds+dus+sdu+usd+sud+dsu)~~\\
			\chi_{1,1}^{f1} &= \sqrt{\frac{1}{6}}(2uus-usu-suu)~~\\ 
			              \chi_{1,1}^{f2}   &= \sqrt{\frac{1}{2}}(usu-suu)~~\\			
			             \chi_{1,1}^{f3}  &= \sqrt{\frac{1}{6}}(2uub-ubu-buu)~~\\
			             \chi_{1,1}^{f4}  &= \sqrt{\frac{1}{2}}(ubu-buu)~~\\			
			              \chi_{1,1}^{f5} &= \sqrt{\frac{1}{3}}(uub+ubu+buu)~~\\ 
			               \chi_{1,1}^{f6}   &= \sqrt{\frac{1}{3}}(uus+usu+suu)~~\\ 
			\end{align*}

Similar to the spin wave function, we use $\chi_{I,I_{z}}^{f}$ to denote the baryon flavor wave function, where $I$ and $I_z$ represent the isospin quantum number and its third component, respectively. Here, both light quarks and heavy quarks are treated as identical particles with $SU\left(5\right)$ symmetry.

For baryon color wave function: Each quark carries one of three color charges: red ($r$), green ($g$), or blue ($b$). As is well known, all physically observed hadronic states must be color singlets (colorless), which implies that the color wave function must be fully antisymmetric. The color wave function for a baryon cluster can be expressed as:	
     \begin{align*}
     \chi^c=\sqrt{\frac{1}{6}}\left(rgb-rbg+gbr-grb+brg-bgr\right)
     \end{align*}
	
	The combined color, spin, and flavor wave function of the baryon is denoted as $\phi_{I_z,S_z}^B$, where $B$ denotes the baryon. The system involves a total of 61 distinct
baryon wave functions:
where the expression of $\phi_{I_{z},S_{z} }^{B}$ are shown as follow:
\begin{align*}
\phi _{0,\frac{1}{2}}^{\Lambda^{0}}&= \sqrt{\frac{1}{2} }\left (  \chi _{0,0}^{f1}\chi _{\frac{1}{2},\frac{1}{2}}^{\sigma 1} +\chi _{0,0}^{f2}\chi _{\frac{1}{2},\frac{1}{2}}^{\sigma 2} \right )\chi ^{c} \\
\phi _{0,\frac{1}{2}}^{\Lambda^{0}_{b}}&= \sqrt{\frac{1}{2}}\left (  \chi _{0,0}^{f3}\chi _{\frac{1}{2},\frac{1}{2}}^{\sigma 1} +\chi _{0,0}^{f4}\chi _{\frac{1}{2},\frac{1}{2}}^{\sigma 2} \right )\chi ^{c}\\
\phi _{\frac{1}{2},\frac{1}{2}}^{p^{+}}&= \sqrt{\frac{1}{2}}\left (  \chi _{\frac{1}{2},\frac{1}{2}}^{f1}\chi _{\frac{1}{2},\frac{1}{2}}^{\sigma 1} +\chi _{\frac{1}{2},\frac{1}{2}}^{f2}\chi _{\frac{1}{2},\frac{1}{2}}^{\sigma 2} \right )\chi ^{c}\\
\phi _{-\frac{1}{2},\frac{1}{2}}^{\Xi^{-}_{b}}&= \sqrt{\frac{1}{2}}\left (  \chi _{\frac{1}{2},-\frac{1}{2}}^{f1}\chi _{\frac{1}{2},\frac{1}{2}}^{\sigma 1} +\chi _{\frac{1}{2},-\frac{1}{2}}^{f2}\chi _{\frac{1}{2},\frac{1}{2}}^{\sigma 2} \right )\chi ^{c}\\
\phi _{-\frac{1}{2},\frac{1}{2}}^{n^{0}}&= \sqrt{\frac{1}{2}}\left (  \chi _{\frac{1}{2},-\frac{1}{2}}^{f3}\chi _{\frac{1}{2},\frac{1}{2}}^{\sigma 1} +\chi _{\frac{1}{2},-\frac{1}{2}}^{f4}\chi _{\frac{1}{2},\frac{1}{2}}^{\sigma 2} \right )\chi ^{c}\\
\phi _{\frac{1}{2},\frac{1}{2}}^{\Xi^{0}_{b}}&= \sqrt{\frac{1}{2}}\left (  \chi _{\frac{1}{2},\frac{1}{2}}^{f3}\chi _{\frac{1}{2},\frac{1}{2}}^{\sigma 1} +\chi _{\frac{1}{2},\frac{1}{2}}^{f4}\chi _{\frac{1}{2},\frac{1}{2}}^{\sigma 2} \right )\chi ^{c} \\
\phi _{-\frac{1}{2},\frac{1}{2}}^{\Xi_{b}^{\prime-}}&= \sqrt{\frac{1}{2}}\left (  \chi _{\frac{1}{2},-\frac{1}{2}}^{f5}\chi _{\frac{1}{2},\frac{1}{2}}^{\sigma 1} +\chi _{\frac{1}{2},-\frac{1}{2}}^{f6}\chi _{\frac{1}{2},\frac{1}{2}}^{\sigma 2} \right )\chi ^{c}\\
\phi _{\frac{1}{2},\frac{1}{2}}^{\Xi_{b}^{\prime0}}&= \sqrt{\frac{1}{2}}\left (  \chi _{\frac{1}{2},\frac{1}{2}}^{f5}\chi _{\frac{1}{2},\frac{1}{2}}^{\sigma 1} +\chi _{\frac{1}{2},\frac{1}{2}}^{f6}\chi _{\frac{1}{2},\frac{1}{2}}^{\sigma 2} \right )\chi ^{c}\\
\phi _{1,\frac{1}{2}}^{\Sigma^{+}}&= \sqrt{\frac{1}{2}}\left (  \chi _{1,1}^{f1}\chi _{\frac{1}{2},\frac{1}{2}}^{\sigma 1} +\chi _{1,1}^{f2}\chi _{\frac{1}{2},\frac{1}{2}}^{\sigma 2} \right )\chi ^{c} \\
\phi _{-1,\frac{1}{2}}^{\Sigma^{-}_{b}}&= \sqrt{\frac{1}{2}}\left (  \chi _{1,-1}^{f1}\chi _{\frac{1}{2},\frac{1}{2}}^{\sigma 1} +\chi _{1,-1}^{f2}\chi _{\frac{1}{2},\frac{1}{2}}^{\sigma 2} \right )\chi ^{c}\\
\phi _{0,\frac{1}{2}}^{\Sigma^{0}}&= \sqrt{\frac{1}{2}}\left (  \chi _{1,0}^{f1}\chi _{\frac{1}{2},\frac{1}{2}}^{\sigma 1} +\chi _{1,0}^{f2}\chi _{\frac{1}{2},\frac{1}{2}}^{\sigma 2} \right )\chi ^{c}\\
\phi _{0,\frac{1}{2}}^{\Sigma^{0}_{b}}&= \sqrt{\frac{1}{2}}\left (  \chi _{1,0}^{f3}\chi _{\frac{1}{2},\frac{1}{2}}^{\sigma 1} +\chi _{1,0}^{f4}\chi _{\frac{1}{2},\frac{1}{2}}^{\sigma 2} \right )\chi ^{c} \\
\phi _{-1,\frac{1}{2}}^{\Sigma^{-}}&= \sqrt{\frac{1}{2}}\left (  \chi _{1,-1}^{f3}\chi _{\frac{1}{2},\frac{1}{2}}^{\sigma 1} +\chi _{1,-1}^{f4}\chi _{\frac{1}{2},\frac{1}{2}}^{\sigma 2} \right )\chi ^{c}\\
\phi _{1,\frac{1}{2}}^{\Sigma_{b}^{+}}&= \sqrt{\frac{1}{2}}\left (  \chi _{1,1}^{f3}\chi _{\frac{1}{2},\frac{1}{2}}^{\sigma 1} +\chi _{1,1}^{f4}\chi _{\frac{1}{2},\frac{1}{2}}^{\sigma 2} \right )\chi ^{c}\\
\phi _{0,-\frac{1}{2}}^{\Sigma^{0}}&= \sqrt{\frac{1}{2}}\left (  \chi _{1,0}^{f1}\chi _{\frac{1}{2},-\frac{1}{2}}^{\sigma 1} +\chi _{1,0}^{f2}\chi _{\frac{1}{2},-\frac{1}{2}}^{\sigma 2} \right )\chi ^{c} \\
			     \end{align*}
			      \begin{align*}
\phi _{1,-\frac{1}{2}}^{\Sigma^{+}}&= \sqrt{\frac{1}{2}}\left (  \chi _{1,1}^{f1}\chi _{\frac{1}{2},-\frac{1}{2}}^{\sigma 1} +\chi _{1,1}^{f2}\chi _{\frac{1}{2},-\frac{1}{2}}^{\sigma 2} \right )\chi ^{c}\\
\phi _{-1,-\frac{1}{2}}^{\Sigma^{-}}&= \sqrt{\frac{1}{2}}\left (  \chi _{1,-1}^{f3}\chi _{\frac{1}{2},-\frac{1}{2}}^{\sigma 1} +\chi _{1,-1}^{f4}\chi _{\frac{1}{2},-\frac{1}{2}}^{\sigma 2} \right )\chi ^{c}\\
\phi _{0,\frac{1}{2}}^{\Omega^{-}_{b}}&= \sqrt{\frac{1}{2} }\left (  \chi _{0,0}^{f5}\chi _{\frac{1}{2},\frac{1}{2}}^{\sigma 1} +\chi _{0,0}^{f6}\chi _{\frac{1}{2},\frac{1}{2}}^{\sigma 2} \right )\chi ^{c} \\
\phi _{0,-\frac{1}{2}}^{\Lambda^{0}}&= \sqrt{\frac{1}{2} }\left (  \chi _{0,0}^{f1}\chi _{\frac{1}{2},-\frac{1}{2}}^{\sigma 1} +\chi _{0,0}^{f2}\chi _{\frac{1}{2},-\frac{1}{2}}^{\sigma 2} \right )\chi ^{c}\\
\phi _{\frac{1}{2},\frac{1}{2}}^{\Xi^{0}}&= \sqrt{\frac{1}{2} }\left (  \chi _{\frac{1}{2},\frac{1}{2}}^{f7}\chi _{\frac{1}{2},\frac{1}{2}}^{\sigma 1} +\chi _{\frac{1}{2},\frac{1}{2}}^{f8}\chi _{\frac{1}{2},\frac{1}{2}}^{\sigma 2} \right )\chi ^{c}\\
\phi _{-\frac{1}{2},\frac{1}{2}}^{\Xi^{-}}&= \sqrt{\frac{1}{2} }\left (  \chi _{\frac{1}{2},-\frac{1}{2}}^{f7}\chi _{\frac{1}{2},\frac{1}{2}}^{\sigma 1} +\chi _{\frac{1}{2},-\frac{1}{2}}^{f8}\chi _{\frac{1}{2},\frac{1}{2}}^{\sigma 2} \right )\chi ^{c} \\
\phi _{-\frac{1}{2},-\frac{1}{2}}^{\Xi^{-}}&= \sqrt{\frac{1}{2} }\left (  \chi _{\frac{1}{2},-\frac{1}{2}}^{f7}\chi _{\frac{1}{2},-\frac{1}{2}}^{\sigma 1} +\chi _{\frac{1}{2},-\frac{1}{2}}^{f8}\chi _{\frac{1}{2},-\frac{1}{2}}^{\sigma 2} \right )\chi ^{c}\\
\phi _{\frac{1}{2},-\frac{1}{2}}^{\Xi^{0}}&= \sqrt{\frac{1}{2} }\left (  \chi _{\frac{1}{2},\frac{1}{2}}^{f7}\chi _{\frac{1}{2},-\frac{1}{2}}^{\sigma 1} +\chi _{\frac{1}{2},\frac{1}{2}}^{f8}\chi _{\frac{1}{2},-\frac{1}{2}}^{\sigma 2} \right )\chi ^{c}\\
\phi _{-\frac{1}{2},-\frac{1}{2}}^{\Xi_{b}^{\prime-}}&= \sqrt{\frac{1}{2} }\left (  \chi _{\frac{1}{2},-\frac{1}{2}}^{f5}\chi _{\frac{1}{2},-\frac{1}{2}}^{\sigma 1} +\chi _{\frac{1}{2},-\frac{1}{2}}^{f6}\chi _{\frac{1}{2},-\frac{1}{2}}^{\sigma 2} \right )\chi ^{c}\\
\phi _{\frac{1}{2},-\frac{1}{2}}^{\Xi_{b}^{\prime0}}&= \sqrt{\frac{1}{2} }\left (  \chi _{\frac{1}{2},\frac{1}{2}}^{f5}\chi _{\frac{1}{2},-\frac{1}{2}}^{\sigma 1} +\chi _{\frac{1}{2},\frac{1}{2}}^{f6}\chi _{\frac{1}{2},-\frac{1}{2}}^{\sigma 2} \right )\chi ^{c}\\
\phi _{0,-\frac{1}{2}}^{\Omega^{-}_{b}}&= \sqrt{\frac{1}{2} }\left (  \chi _{0,0}^{f5}\chi _{\frac{1}{2},-\frac{1}{2}}^{\sigma 1} +\chi _{0,0}^{f6}\chi _{\frac{1}{2},-\frac{1}{2}}^{\sigma 2} \right )\chi ^{c} \\
 \end{align*}
			      \begin{align*}
\phi_{0,\frac{3}{2}}^{\Sigma_{b}^{*0}}&=\chi_{1,0}^{f5}\chi_
{\frac{3}{2},\frac{3}{2}}^{\sigma}\chi^{c} &
\phi_{-1,\frac{3}{2}}^{\Sigma_{b}^{*-}}&=\chi_{1,-1}^{f5}
\chi_{\frac{3}{2},\frac{3}{2}}^{\sigma}\chi^{c}\\
\phi_{1,\frac{3}{2}}^{\Sigma_{b}^{*+}}&=\chi_{1,1}^{f5}\chi_
{\frac{3}{2},\frac{3}{2}}^{\sigma}\chi^{c} &
\phi_{0,\frac{1}{2}}^{\Sigma_{b}^{*0}}&=\chi_{1,0}^{f5}
\chi_{\frac{3}{2},\frac{1}{2}}^{\sigma}\chi^{c}  \\
\phi_{-1,\frac{1}{2}}^{\Sigma_{b}^{*-}}&=\chi_{1,-1}^{f5}\chi_
{\frac{3}{2},\frac{1}{2}}^{\sigma}\chi^{c}&
\phi_{1,\frac{1}{2}}^{\Sigma_{b}^{*+}}&=\chi_{1,1}^{f5}
\chi_{\frac{3}{2},\frac{1}{2}}^{\sigma}\chi^{c} \\
\phi_{1,\frac{3}{2}}^{\Sigma^{*+}}&=\chi_{1,1}^{f6}\chi_
{\frac{3}{2},\frac{3}{2}}^{\sigma}\chi^{c}&
\phi_{0,\frac{3}{2}}^{\Sigma^{*0}}&=\chi_{1,0}^{f6}
\chi_{\frac{3}{2},\frac{3}{2}}^{\sigma}\chi^{c} \\
\phi_{1,\frac{1}{2}}^{\Sigma^{*+}}&=\chi_{1,1}^{f6}
\chi_{\frac{3}{2},\frac{1}{2}}^{\sigma}\chi^{c}&
\phi_{0,\frac{1}{2}}^{\Sigma^{*0}}&=\chi_{1,0}^{f6}\chi_
{\frac{3}{2},\frac{1}{2}}^{\sigma}\chi^{c}\\
\phi_{-1,\frac{1}{2}}^{\Sigma^{*-}}&=\chi_{1,-1}^{f6}
\chi_{\frac{3}{2},\frac{1}{2}}^{\sigma}\chi^{c} &
\phi_{-1,-\frac{1}{2}}^{\Sigma_{b}^{*-}}&=\chi_{1,-1}^{f5}\chi_
{\frac{3}{2},-\frac{1}{2}}^{\sigma}\chi^{c}\\
\phi_{0,-\frac{1}{2}}^{\Sigma_{b}^{*0}}&=\chi_{1,0}^{f5}
\chi_{\frac{3}{2},-\frac{1}{2}}^{\sigma}\chi^{c}&
\phi_{1,-\frac{1}{2}}^{\Sigma_{b}^{*+}}&=\chi_{1,1}^{f5}\chi_
{\frac{3}{2},-\frac{1}{2}}^{\sigma}\chi^{c}\\
\phi_{1,-\frac{1}{2}}^{\Sigma^{*+}}&=\chi_{1,1}^{f6}
\chi_{\frac{3}{2},-\frac{1}{2}}^{\sigma}\chi^{c}&
\phi_{0,-\frac{1}{2}}^{\Sigma_{b}^{*0}}&=\chi_{1,0}^{f6}\chi_
{\frac{3}{2},-\frac{1}{2}}^{\sigma}\chi^{c} \\
\phi_{-1,-\frac{1}{2}}^{\Sigma^{*-}}&=\chi_{1,-1}^{f6}
\chi_{\frac{3}{2},-\frac{1}{2}}^{\sigma}\chi^{c}&
\phi_{0,\frac{3}{2}}^{\Omega_{b}^{*-}}&=\chi_{0,0}^{f7}\chi_
{\frac{3}{2},\frac{3}{2}}^{\sigma}\chi^{c}\\
\phi_{0,\frac{1}{2}}^{\Omega_{b}^{*-}}&=\chi_{0,0}^{f7}
\chi_{\frac{3}{2},\frac{1}{2}}^{\sigma}\chi^{c}&
\phi_{0,\frac{1}{2}}^{\Omega^{*-}}&=\chi_{0,0}^{f8}\chi_
{\frac{3}{2},\frac{1}{2}}^{\sigma}\chi^{c}\\
\phi_{0,\frac{3}{2}}^{\Omega^{*-}}&=\chi_{0,0}^{f8}
\chi_{\frac{3}{2},\frac{3}{2}}^{\sigma}\chi^{c} &
\phi_{\frac{1}{2},\frac{3}{2}}^{\Xi_{b}^{*0}}&=\chi_
{\frac{1}{2},\frac{1}{2}}^{f9}\chi_{\frac{3}{2},
\frac{3}{2}}^{\sigma}\chi^{c}\\
\phi_{-\frac{1}{2},\frac{3}{2}}^{\Xi_{b}^{*-}}&=\chi_
{\frac{1}{2},-\frac{1}{2}}^{f9}\chi_{\frac{3}{2},
\frac{3}{2}}^{\sigma}\chi^{c}&
\phi_{\frac{1}{2},\frac{1}{2}}^{\Xi_{b}^{*0}}&=\chi_
{\frac{1}{2},\frac{1}{2}}^{f9}\chi_{\frac{3}{2},
\frac{1}{2}}^{\sigma}\chi^{c} \\
\phi_{-\frac{1}{2},\frac{1}{2}}^{\Xi_{b}^{*-}}&=\chi_
{\frac{1}{2},-\frac{1}{2}}^{f9}\chi_{\frac{3}{2},
\frac{1}{2}}^{\sigma}\chi^{c}&
\phi_{\frac{1}{2},\frac{3}{2}}^{\Xi^{*0}}&=\chi_
{\frac{1}{2},\frac{1}{2}}^{f10}\chi_{\frac{3}{2},
\frac{3}{2}}^{\sigma}\chi^{c}\\
\phi_{-\frac{1}{2},\frac{3}{2}}^{\Xi^{*-}}&=\chi_
{\frac{1}{2},-\frac{1}{2}}^{f10}\chi_{\frac{3}{2},
\frac{3}{2}}^{\sigma}\chi^{c}&
\phi_{\frac{1}{2},\frac{1}{2}}^{\Xi^{*0}}&=\chi_
{\frac{1}{2},\frac{1}{2}}^{f10}\chi_{\frac{3}{2},
\frac{1}{2}}^{\sigma}\chi^{c} \\
			     \end{align*}
			      \begin{align*}
\phi_{-\frac{1}{2},\frac{1}{2}}^{\Xi^{*-}}&=\chi_
{\frac{1}{2},-\frac{1}{2}}^{f10}\chi_{\frac{3}{2},
\frac{1}{2}}^{\sigma}\chi^{c}&
\phi_{-\frac{1}{2},-\frac{1}{2}}^{\Xi_{b}^{*-}}&=\chi_
{\frac{1}{2},-\frac{1}{2}}^{f9}\chi_{\frac{3}{2},
-\frac{1}{2}}^{\sigma}\chi^{c}\\
\phi_{\frac{1}{2},-\frac{1}{2}}^{\Xi_{b}^{*0}}&=\chi_
{\frac{1}{2},\frac{1}{2}}^{f9}\chi_{\frac{3}{2},
-\frac{1}{2}}^{\sigma}\chi^{c}&
\phi_{\frac{1}{2},-\frac{1}{2}}^{\Xi^{*0}}&=\chi_
{\frac{1}{2},\frac{1}{2}}^{f10}\chi_{\frac{3}{2},
-\frac{1}{2}}^{\sigma}\chi^{c} \\
\phi_{-\frac{1}{2},-\frac{1}{2}}^{\Xi^{*-}}&=\chi_
{\frac{1}{2},-\frac{1}{2}}^{f10}\chi_{\frac{3}{2},
-\frac{1}{2}}^{\sigma}\chi^{c}&
\phi_{0,-\frac{1}{2}}^{\Omega^{*-}}&=\chi_{0,0}^{f8}\chi_
{\frac{3}{2},-\frac{1}{2}}^{\sigma}\chi^{c}\\
\phi_{0,\frac{3}{2}}^{\Omega_{b}^{*-}}&=\chi_{0,0}^{f7}
\chi_{\frac{3}{2},-\frac{1}{2}}^{\sigma}\chi^{c} &
\end{align*}

By substituting the wave functions of the flavor, spin, and color components according to the given quantum numbers of the system, the total flavor-spin-color wave function of the dibaryon system can be obtained. The total wave functions are presented as follow:

For the $B=1,~S=-1$ system:
		\begin{align*}
			\left | \Lambda \Lambda _{b}   \right \rangle &= \phi _{0,\frac{1}{2} }^{\Lambda^{0 } } \phi _{0,\frac{1}{2} }^{\Lambda^{0 } _{b}}~~    \\
			\left | N\Xi _{b}  \right \rangle &=\sqrt{\frac{1}{2} } \left [ \phi _{\frac{1}{2},\frac{1}{2}  }^{p^{+ }} \phi _{-\frac{1}{2} ,\frac{1}{2} }^{\Xi^{- } _{b} } -\phi _{-\frac{1}{2},\frac{1}{2}  }^{n^{0 }} \phi _{\frac{1}{2} ,\frac{1}{2} }^{\Xi^{0 } _{b} }\right ]~~    \\
			\left | N\Xi _{b}^{\prime }  \right \rangle &=\sqrt{\frac{1}{2} } \left [ \phi _{\frac{1}{2},\frac{1}{2}  }^{p^{+ }} \phi _{-\frac{1}{2} ,\frac{1}{2} }^{\Xi _{b} ^{\prime-} } -\phi _{-\frac{1}{2},\frac{1}{2}  }^{n^{0 }} \phi _{\frac{1}{2} ,\frac{1}{2} }^{\Xi _{b}^{\prime0 } }\right ]~~    \\
			\left | N\Xi _{b}^{\ast }  \right \rangle&=\sqrt{\frac{3}{8} }  \phi _{\frac{1}{2},\frac{3}{2}  }^{\Xi _{b}^{\ast 0}} \phi _{-\frac{1}{2} ,-\frac{1}{2} }^{n ^{0 }} -\sqrt{\frac{1}{8} }\phi _{\frac{1}{2},\frac{1}{2}  }^{\Xi _{b}^{\ast 0 }} \phi _{-\frac{1}{2} ,\frac{1}{2} }^{n^{0 } } \\
			  &-\sqrt{\frac{3}{8} }\phi _{-\frac{1}{2},\frac{3}{2}  }^{\Xi _{b}^{\ast -}} \phi _{\frac{1}{2} ,-\frac{1}{2} }^{p ^{+ }}+\sqrt{\frac{1}{8} }\phi _{-\frac{1}{2},\frac{1}{2}  }^{\Xi _{b}^{\ast -}} \phi _{\frac{1}{2} ,\frac{1}{2} }^{p ^{+ }}~~   \\
		\left | \Sigma \Sigma _{b} \right \rangle &=\sqrt{\frac{1}{3} } \left [ \phi _{1,\frac{1}{2} }^{\Sigma^{+ }}\phi _{-1,\frac{1}{2} }^{\Sigma^{- } _{b} }-\phi _{0,\frac{1}{2} }^{\Sigma^{0 }}\phi _{0,\frac{1}{2} }^{\Sigma ^{0 }_{b} }  +\phi _{-1,\frac{1}{2} }^{\Sigma^{- }}\phi _{1,\frac{1}{2} }^{\Sigma^{+ } _{b} } \right ]~~      \\	
			\left |\Sigma \Sigma _{b}^{*}  \right \rangle &=\frac{1}{2} \left [ \phi _{0,-\frac{1}{2} }^{\Sigma ^{0 }}\phi _{0,\frac{3}{2} }^{\Sigma _{b}^{*0}}-\phi _{1,-\frac{1}{2} }^{\Sigma^{+ } }\phi _{-1,\frac{3}{2} }^{\Sigma _{b}^{*-}}-\phi _{-1,-\frac{1}{2} }^{\Sigma ^{- }}\phi _{1,\frac{3}{2} }^{\Sigma _{b}^{*+}}\right ]~~     \\
			& -\sqrt{\frac{1}{12}} \left [ \phi _{0,\frac{1}{2} }^{\Sigma ^{0 }}\phi _{0,\frac{1}{2} }^{\Sigma _{b}^{*0}}- \phi _{1,\frac{1}{2} }^{\Sigma ^{+ }}\phi _{-1,\frac{1}{2} }^{\Sigma _{b}^{*-}}-\phi _{-1,\frac{1}{2} }^{\Sigma^{- } }\phi _{1,\frac{1}{2} }^{\Sigma _{b}^{*+}}\right ] \\
			\left |\Sigma^{*}  \Sigma _{b} \right \rangle &= \frac{1}{2} \left [ \phi _{1,\frac{3}{2} }^{\Sigma^{*+} }\phi _{-1,-\frac{1}{2} }^{\Sigma^{- } _{b}}-\phi _{0,\frac{3}{2} }^{\Sigma^{*0} }\phi _{0,-\frac{1}{2} }^{\Sigma ^{0 }_{b}}+\phi _{-1,\frac{3}{2} }^{\Sigma^{*-} }\phi _{1,-\frac{1}{2} }^{\Sigma^{+ } _{b}}\right ]~~     \\
			& -\sqrt{\frac{1}{12}} \left [ \phi _{1,\frac{1}{2} }^{\Sigma ^{*+}}\phi _{-1,\frac{1}{2} }^{\Sigma ^{-}_{b}}- \phi _{0,\frac{1}{2} }^{\Sigma ^{*0}}\phi _{0,\frac{1}{2} }^{\Sigma^{0 } _{b}}-\phi _{-1,\frac{1}{2} }^{\Sigma^{*-} }\phi _{1,\frac{1}{2} }^{\Sigma^{+ } _{b}}\right ] \\
			\left |\Sigma^{*}  \Sigma _{b}^{*} \right \rangle &= \sqrt{\frac{1}{10} }  \left [ \phi _{1,\frac{3}{2} }^{\Sigma^{*+} }\phi _{-1,-\frac{1}{2} }^{\Sigma _{b}^{*-}}-\phi _{0,\frac{3}{2} }^{\Sigma^{*0} }\phi _{0,-\frac{1}{2} }^{\Sigma _{b}^{*0}}+\phi _{-1,\frac{3}{2} }^{\Sigma^{*-} }\phi _{1,-\frac{1}{2} }^{\Sigma _{b}^{*+}}\right]~~   \\
			&+\sqrt{\frac{1}{10} }\left [\phi _{1,-\frac{1}{2} }^{\Sigma^{*+} }\phi _{-1,\frac{3}{2} }^{\Sigma _{b}^{*-}}-\phi _{0,-\frac{1}{2} }^{\Sigma^{*0} }\phi _{0,\frac{3}{2} }^{\Sigma _{b}^{*0}}+\phi _{-1,-\frac{1}{2} }^{\Sigma^{*-} }\phi _{1,\frac{3}{2} }^{\Sigma _{b}^{*+}}\right ]\\
			&-\sqrt{\frac{2}{15}} \left [ \phi _{1,\frac{1}{2} }^{\Sigma ^{*+}}\phi _{-1,\frac{1}{2} }^{\Sigma _{b}^{*-}}- \phi _{0,\frac{1}{2} }^{\Sigma ^{*0}}\phi _{0,\frac{1}{2} }^{\Sigma _{b}^{*0}}+\phi _{-1,\frac{1}{2} }^{\Sigma^{*-} }\phi _{1,\frac{1}{2} }^{\Sigma _{b}^{*+}}\right ]\\
			\end{align*}
For the $B=1,~S=-3$ system:
\begin{align*}
			\left | \Lambda \Omega _{b}\right \rangle &= \phi _{0,\frac{1}{2} }^{\Lambda^{0 } } \phi _{0,\frac{1}{2} }^{\Omega^{-} _{b}} \\			
			\left | \Lambda \Omega _{b}^{*} \right \rangle &=\frac{1}{2}\phi _{0,\frac{1}{2} }^{\Lambda^{0 } } \phi _{0,\frac{1}{2} }^{\Omega _{b}^{*-}}-\sqrt{\frac{3}{4} }\phi _{0,-\frac{1}{2} }^{\Lambda ^{0 }} \phi _{0,\frac{3}{2} }^{\Omega _{b}^{*-}} ~~    \\			
			\left | \Lambda _{b}\Omega \right \rangle &=\frac{1}{2}\phi _{0,\frac{1}{2} }^{\Lambda ^{0 }_{b}} \phi _{0,\frac{1}{2} }^{\Omega^{- } }-\sqrt{\frac{3}{4} }\phi _{0,-\frac{1}{2} }^{\Lambda ^{0 }_{b}} \phi _{0,\frac{3}{2} }^{\Omega^{- } }~~    \\	
						     \end{align*}
			      \begin{align*}		
			\left |\Xi \Xi _{b}  \right \rangle &= \sqrt{\frac{1}{2} } \left [ \phi _{\frac{1}{2},\frac{1}{2}  }^{\Xi^{0 } } \phi _{-\frac{1}{2},\frac{1}{2}  }^{\Xi ^{- }_{b}} - \phi _{-\frac{1}{2},\frac{1}{2}  }^{\Xi^{- } } \phi _{\frac{1}{2},\frac{1}{2}}^{\Xi^{0 } _{b}} \right ]~~   \\		
			\left |\Xi \Xi _{b}^{\prime }  \right \rangle &=\sqrt{\frac{1}{2} } \left [ \phi _{\frac{1}{2},\frac{1}{2}  }^{\Xi^{0 } } \phi _{-\frac{1}{2},\frac{1}{2}  }^{\Xi _{b}^{\prime- }} - \phi _{-\frac{1}{2},\frac{1}{2}  }^{\Xi ^{- }} \phi _{\frac{1}{2},\frac{1}{2}  }^{\Xi _{b}^{\prime 0}} \right ]~~      \\			
			\left |\Xi \Xi _{b}^{*}\right \rangle &=\sqrt{\frac{3}{8}}\left [ \phi _{-\frac{1}{2},-\frac{1}{2}  }^{\Xi^{- } } \phi _{\frac{1}{2},\frac{3}{2}  }^{\Xi _{b}^{*0}} - \phi _{\frac{1}{2},-\frac{1}{2}  }^{\Xi^{0 } } \phi _{-\frac{1}{2},\frac{3}{2}  }^{\Xi _{b}^{*- }} \right ]~~     \\
			&-\sqrt{\frac{1}{8} }\left [\phi _{-\frac{1}{2},\frac{1}{2}  }^{\Xi^{- } } \phi _{\frac{1}{2},\frac{1}{2}  }^{\Xi _{b}^{*0 }} - \phi _{\frac{1}{2},\frac{1}{2}  }^{\Xi ^{0 }} \phi _{-\frac{1}{2},\frac{1}{2}  }^{\Xi _{b}^{* -}}  \right ] \\			
			\left |\Xi^{*} \Xi _{b}\right \rangle &=\sqrt{\frac{3}{8}}\left [ \phi _{\frac{1}{2},\frac{3}{2}  }^{\Xi ^{* 0}} \phi _{-\frac{1}{2},-\frac{1}{2}  }^{\Xi ^{- }_{b}} - \phi _{-\frac{1}{2},\frac{3}{2}  }^{\Xi ^{*- }} \phi _{\frac{1}{2},-\frac{1}{2}  }^{\Xi^{0 } _{b}} \right ] ~~     \\
			&-\sqrt{\frac{1}{8} }\left [\phi _{\frac{1}{2},\frac{1}{2}  }^{\Xi^{* 0} } \phi _{-\frac{1}{2},\frac{1}{2}  }^{\Xi ^{- }_{b}} - \phi _{-\frac{1}{2},\frac{1}{2}  }^{\Xi ^{*- }} \phi _{\frac{1}{2},\frac{1}{2}  }^{\Xi^{0 }_{b}}  \right ]\\		
			\left |\Xi^{*} \Xi _{b}^{\prime }\right \rangle &=   \sqrt{\frac{3}{8}}\left [ \phi _{\frac{1}{2},\frac{3}{2}  }^{\Xi ^{* 0}} \phi _{-\frac{1}{2},-\frac{1}{2}  }^{\Xi _{b}^{\prime- }} - \phi _{-\frac{1}{2},\frac{3}{2}  }^{\Xi ^{*-}} \phi _{\frac{1}{2},-\frac{1}{2}  }^{\Xi _{b}^{\prime 0}} \right ] ~~   \\
			&-\sqrt{\frac{1}{8} }\left [\phi _{\frac{1}{2},\frac{1}{2}  }^{\Xi^{*0 } } \phi _{-\frac{1}{2},\frac{1}{2}  }^{\Xi _{b}^{\prime -}} - \phi _{-\frac{1}{2},\frac{1}{2}  }^{\Xi ^{*- }} \phi _{\frac{1}{2},\frac{1}{2}  }^{\Xi _{b}^{\prime0 }}  \right ]  \\
			\end{align*}
			      \begin{align*}
			\left |\Xi^{*} \Xi _{b}^{*}\right \rangle &= \sqrt{\frac{3}{20}}\left [ \phi _{\frac{1}{2},\frac{3}{2}  }^{\Xi ^{* 0}} \phi _{-\frac{1}{2},-\frac{1}{2}  }^{\Xi _{b}^{*-}} - \phi _{-\frac{1}{2},\frac{3}{2}  }^{\Xi ^{*- }} \phi _{\frac{1}{2},-\frac{1}{2}  }^{\Xi _{b}^{*0}}\right ]\\
			&+\sqrt{\frac{3}{20}}\left [\phi _{\frac{1}{2},-\frac{1}{2}  }^{\Xi ^{* 0}} \phi _{-\frac{1}{2},\frac{3}{2}  }^{\Xi _{b}^{*-}}-\phi _{-\frac{1}{2},-\frac{1}{2}  }^{\Xi ^{*-}} \phi _{\frac{1}{2},\frac{3}{2}  }^{\Xi _{b}^{*0}}\right ] \\
			&-\sqrt{\frac{3}{10} }\left [\phi _{\frac{1}{2},\frac{1}{2}  }^{\Xi ^{*0}} \phi _{-\frac{1}{2},\frac{1}{2}  }^{\Xi _{b}^{*-}}-\phi _{-\frac{1}{2},\frac{1}{2}  }^{\Xi ^{* -}} \phi _{\frac{1}{2},\frac{1}{2}  }^{\Xi _{b}^{*0}}\right ] \\
			\end{align*}
For the $B=1,~S=-5$ system:
\begin{align*}
			\left | \Omega \Omega _{b}\right \rangle &=-\sqrt{\frac{3}{4} } \phi _{0,-\frac{1}{2} }^{\Omega^{- }_{b}}\phi _{0,\frac{3}{2} }^{\Omega^{- }}+\sqrt{\frac{1}{4} }\phi _{0,\frac{1}{2} }^{\Omega^{- }_{b}}\phi _{0,\frac{1}{2} }^{\Omega^{- }}~~    \\
			\left | \Omega \Omega _{b}^{*}  \right \rangle &= \sqrt{\frac{3}{10} }\left [ \phi _{0,\frac{3}{2} }^{\Omega^{- }} \phi _{0,-\frac{1}{2} }^{\Omega_{b}^{*-} } +\phi _{0,-\frac{1}{2} }^{\Omega^{- }} \phi _{0,\frac{3}{2} }^{\Omega_{b}^{*-} }\right ]~~    \\
			&-\sqrt{\frac{2}{5} }\phi _{0,\frac{1}{2}}^{\Omega^{- } }\phi _{0,\frac{1}{2} }^{\Omega _{b} ^{*-}}\\
			\end{align*}

\end{document}